\documentclass[aps,prd,onecolumn,amssymb]{revtex4}
\usepackage{graphicx,bm,color}
\usepackage{amsmath}
\usepackage{amssymb}
\usepackage{amsfonts}
\usepackage{natbib}
\usepackage{inputenc}
\newcommand{\be}{\begin{equation}}
\newcommand{\ee}{\end{equation}}
\newcommand{\bea}{\begin{eqnarray}}
\newcommand{\eea}{\end{eqnarray}}
\newcommand{\beaa}{\begin{eqnarray*}}
\newcommand{\eeaa}{\end{eqnarray*}}

\newcommand{\nn}{\nonumber \\}

\allowdisplaybreaks[4]

\begin{document}

\title{Cosmological future singularities in massive gravity and massive bigravity}
\author{M. Mousavi}\email{ mousavi@azaruniv.ac.ir}
\author{K. Atazadeh}\email{ atazadeh@azaruniv.ac.ir}

\affiliation{Department of Physics, Azarbaijan Shahid Madani University, Tabriz, 53714-161 Iran}

\begin{abstract}
We study the future cosmological singularities in the framework of massive gravity and minimal massive bigravity theory. In this regards, we consider the possible classes of finite-time future singularities such as sudden, big rip, big freeze and big brake singularities in the massive universe. In dRGT model with an open expanding universe we obtain the sudden singularity in the future at a finite-time which generally without taking account of any particular realistic equation of state, is not avoidable and except the fluid density, all dynamical physical quantities such as pressure approach to infinity. To complete our study, we search the future cosmological singularities in the context of minimal massive bigravity theory and we find that the cosmology of this theory suffers from the sudden and big brake singularities, in which we can see that the parameters of the model approaches to zero the sudden singularity can be removed.
\end{abstract}


\maketitle

\section{Introduction \label{Sec1}}

The discovery of current cosmic acceleration via the type Ia supernovas data \cite{1,2}, the cosmic microwave background (CMB) radiation \cite{3} and the large scale structure\cite{4,5}, has constructed a renewed interest in theories that modify standard gravity. Apart from current explanations as the cosmological constant scenario and the dynamical dark energy option, there is also  the belief that the theory of general relativity (GR) is no longer working on the cosmological scale and needs a modification. As a result, many modified theories have been introduced to justify the present expanding acceleration of our universe without the need of introducing the unknown dark energy. Giving a tiny mass to graviton is an alternative modification of general relativity which can cause a small cosmological constant term to cope with the rationalizing this picture of accelerated expanding universe. About the old history of massive gravity as a comparably successful modified theory we should refer to the year 1939, when Fierz and Pauli proposed their linear model \cite{6}. Actually, the Fierz-Pauli linear action dose not eventuate the linearized Einstein gravity action in the zero-mass limit besides cannot satisfy the solar system tests resulted of the vDVZ discontinuity \cite{7}. The familiar Vainshtein mechanism says that nonlinear interactions can avoid this discontinuity \cite{8} and successively, the nonlinear terms create a Boulware-Deser ghost \cite{9,10}. In the end, a ghost-free nonlinear massive gravity theory was prepared successfully by de Rham, Gabadadze and Tolly (dRGT) \cite{11}. In this manner, by studding \cite{12} we can review the levels leading to the modern approach.\\
In the scenario of massive gravity, in addition to the principal metric $g_{\mu\nu}$ a second tensor field $f_{\mu\nu}$ plays a key role. In this respect, the dRGT theory was presented as a ghost-free one \cite{13}. It is notable that to say the cosmology in this model dose not support the flat Friedmann-Robertson-Walker (FRW) solution \cite{14}, thus it may be ruled out by cosmological observations. It is noticeable that open cosmological solutions have been obtained with two classifications of solutions in which the massive terms in the field equations lead to an effective cosmological constant \cite{15}.
Additionally, about the birth of the massive bigravity theory as a modified massive gravity theory one can refer to \cite{16,17}, in which the authors convert the second tensor field to a dynamical one, similarly as principal metric $g_{\mu\nu}$, although the only difference in this work is that the matter source just couples to the auxiliary metric. Thus, the resulted ghost-free massive bigravity theory allows cosmologically viable solutions. As discussed, two dynamical metrics by which we construct this theory behave in a completely symmetric way and then breaks the belief in aether-like concept of the principal metric in massive gravity. These mentioned features of massive gravity and massive bigravity models encourages people to go more through the deep study of current cosmological issues in them. In the several literatures, the cosmological solutions in massive gravity and massive bigravity theories have considered in Refs. \cite{18,19,20,21,22,23}.\\
Actually, the discovery of the accelerated expanding universe creates much enthusiasms to do research about the features of the cosmic fluid which causes this accelerating expansion named dark energy picturing the overabundance of new types of singularities which are completely different than the initial big bang singularity, the so called ``finite-time future singularities''. Empathically, about the initial cosmological singularity in the early universe we are supposed to refer to the emergent universe scenario in which the universe emerges from a static state as the Einstein static universe (ESU) indicating a non-zero spatial volume with a positive curvature by which the initial big bang singularity will be avoided \cite{24}. Around our discussion, the ESU in massive gravity model has been studied in \cite{25}, in which we see that all curvature spaces $\kappa=\pm1$ and $\kappa=0$ are allowed to describe the ESU.
Also, the initial singularity problem and the ESU in the  massive bigravity models have been considered in \cite{26}, in which the authors have obtained that the flat FRW universe is not capable to explain the ESU. As a result in the following study we take $\kappa=-1$ and choose an open universe.

Many works have studied about the theoretical possibility that expanding universes end up a violent final destiny at a future finite-time without meeting an expansion maximum or sequent collapse to a ``big crunch'' singularity. These new types of singularities are characterized by the violation of some of the energy conditions. As a result, we see that this violation causes divergencies in some of the physical quantities involving the scale factor, energy density and pressure profile. As examples, one may mention big rip singularity \cite{27}, sudden singularity \cite{28}, big freeze singularity \cite{29} and other classifications for the possible future singularities \cite{30}. Additionally, there are some works about cosmological singularity issues which have been studied beyond the Einstein's general relativity \cite{30r,300}. Thus, regarding the interesting aspects of the massive gravity and massive bigravity theories, we have a great motivation to study the features of the future cosmological singularities in these modified self accelerating gravity models.

The paper is organized as follows. In section II, we review briefly the massive gravity theory and its cosmological equations. In section III, we study the possible finite-time future singularities in the massive universe. In section IV, we consider the massive bigravity theory with its modified Friedmann equations and then we address the possible future finite-time cosmological singularities in massive bigravity theory, in section V. Finally, we close the paper whit a conclusion in section VI.

\section{Friedman equations in the Massive Gravity Theory \label{Sec2}}

The theory of massive gravity is introduced by the following action as \cite{14}

\be
\label{Fbi1}
S=M_{g}^{2}\int d^{4}x\sqrt{- {\rm det} g}\left(-\frac{1}{2}R+m^{2}\mathcal{L}_{Int}\right)+S_{m},
\ee

where $M_{g}$, $R$ and $m$ are the Planck mass of the metric $g_{\mu\nu}$, Ricci scalar and the graviton mass, respectively. We should note that $S_{m}$ is just describing the ordinary matter and $\mathcal{L}_{Int}$, non-derivative massive graviton terms defined as

\be
\label{Fbi2}
\mathcal{L}_{Int}=\frac{1}{2}\left(S^{2}-S^{A}_{B}S^{B}_{A}\right)+\frac{c_{3}}{3!}\epsilon_{MNPQ}\epsilon^{ABCQ}S^{M}_{A}S^{N}_{B}S^{P}_{C}
+\frac{c_{4}}{4!}\epsilon_{MNPQ}\epsilon^{ABCQ}S^{M}_{A}S^{N}_{B}S^{Q}_{D},
\ee

that we have $S=S^{A}_{A}$ and also $c_{3}$ and $c_{4}$ are two constants beside that $\epsilon_{MNPQ}$ is the Levi-Civita tensor density. Moreover

\be
\label{Fbi3}
S^{A}_{B}=\delta^{A}_{B}-\gamma^{A}_{B},
\ee
in which $\gamma^{A}_{B}$ is defined by

\be
\label{Fbi4}
\gamma^{A}_{B}\gamma^{C}_{A}=g^{AC}f_{AB}.
\ee

Here $f_{AB}$ is a symmetric tensor field. The original dRGT theory respects the Poincar\`{e} symmetry in the field space and thus the reference metric is Minkowski, {\it i.e.} $f_{AB}=\eta_{AB}$$=$ diag$(-1, 1, 1, 1)$.\\
The spatially homogeneous and isotropic FRW metric can be written as
\be
\label{Fbi5}
ds^{2}=-dt^{2}+a^{2}(t)\left(\frac{dr^{2}}{1-\kappa r^{2}}+r^{2}d^{2}\Omega\right),
\ee
here $a(t)$ is the cosmic scale factor, $t$ is the cosmic time and $\kappa=0,\pm 1$ plays the role of the three dimensional space constant curvature. According to \cite{31} the first Friedmann equation with dimensionless constants $c_{3}$ and $c_{4}$ by imposing the Bianchi constraints is as follows

\be
\label{Fbi6}
3\frac{\dot{a}^{2}+\kappa}{a^{2}}=m^{2}\left(4c_{3}+c_{4}-6+3\mu\frac{3-3c_{3}-c_{4}}{a}+
3\mu^{2}\frac{c_{4}+2c_{3}-1}{a^{2}}-\mu^{3}\frac{c_{3}+c_{4}}{a^{3}}\right)+\frac{\rho}{M_{g}^{2}}.
\ee

Where $\mu$ is an integrating constant. As a result of minimal matter coupling to gravity, the matter equation of motion reads

\be
\label{Fbi7}
\dot{\rho}+3H\left(\rho+P\right)=0,
\ee

that $H=\frac{\dot{a}}{a}$. If we assume an equation of state parameter, namely $\omega$ then we have $P=\omega \rho$. According to \cite{32} the simplest choice that presents a successful Vainshtein effect in the weak field limit is reducing the parameter space to the subset $c_{3}=-c_{4}$. Using this simplification, the first and second Friedmann equations are extracted as follows

\be
\label{Fbi8}
H^{2}=\frac{\rho}{3M_{g}^{2}}-\frac{\kappa}{a^{2}}+\frac{m^{2}}{3}\left(h_{1}+\frac{h_{2}}{a}+\frac{h_{3}}{a^{2}}\right),
\ee

\be
\label{Fbi9}
\dot{H}=\frac{-\rho}{2M_{g}^{2}} \left(1+\omega\right)+\frac{\kappa}{a^{2}}-\frac{m^{2}}{6}\left(\frac{h_{2}}{a}+\frac{2h_{3}}{a^{2}}\right).
\ee
Where $\frac{a}{\mu}\rightarrow a$ and also

\begin{align}
\label{Fbi10}
h_{1}=&-3c_{4}-6,\nn
h_{2}=&3\left(3+2c_{4}\right),\nn
h_{3}=&-3\left(1+c_{4}\right).\nn
\end{align}

As we have $\dot{H}=\frac{\ddot{a}}{a}-\frac{\dot{a}^{2}}{a^{2}}$, thus we can write

\be
\label{Fbi11}
\frac{\ddot{a}}{a}-\frac{\dot{a}^{2}}{a^{2}}+\frac{\rho}{2M_{g}^{2}} \left(1+\omega\right)-\frac{\kappa}{a^{2}}+\frac{m^{2}}{6}\left(\frac{h_{2}}{a}+\frac{2h_{3}}{a^{2}}\right)=0.
\ee

Rearranging the firs Friedmann equation (\ref{Fbi8}) and also the above equation, we have

\be
\label{Fbi12}
\frac{\rho}{3M_{g}^{2}}-\frac{\dot{a}^{2}}{a^{2}}-
\frac{\kappa}{a^{2}}+
\frac{m^{2}}{3}
\left(-3c_{4}
-6 +
\frac{1}{a}\left(3\left(3+2c_{4}\right)\right)+
\frac{1}{a^{2}}\left(-3\left(1+c_{4}\right)\right)\right)=0,
\ee

and

\be
\label{Fbi13}
 \frac{\rho }{2M_{g}^{2}}\left(1+\omega\right)-\frac{\kappa}{a^{2}}+\frac{\ddot{a}}{a}-\frac{\dot{^{2}}}{a^{2}}+\frac{m^{2}}{6}
\left(\frac{1}{a}\left(3\left(3+2c_{4}\right)\right)+\frac{1}{a^{2}}\left(-6\left(1+c_{4}\right)\right)\right)=0.
\ee

One finds $\rho(t)$ and $P(t)$ terms

\be
\label{Fbi14}
\rho(t)=\frac{3M_{g}^{2}}{a^{2}} \left(\dot{a}^{2}+\kappa+m^{2}\left(\left(c_{4}+2\right)a^{2}-\left(3+2c_{4}\right)a+1+c_{4}\right)\right).
\ee

\be
\label{Fbi15}
P(t)=-\frac{M_{g}^{2}}{ a^{2}}\left(2\ddot{a}a+\dot{a}^{2}+\kappa+m^{2}\left(3\left(2+c_{4}\right)a^{2}-2
\left(3+2c_{4}\right)a+1+c_{4}\right)\right).
\ee

In the next section we try to study the possible types of finite-time singularities may be occurred in the future of an open FRW universe in the context of massive gravity theory.\\

\section{Future cosmological singularities in Massive Gravity Theory\label{Sec3}}

As mentioned in the introduction the possible types of finite-time future singularities in GR have been studied by Barrow \cite{28,34}. We are also interested in analyzing massive gravity theory in this research area to obtain all allowed types of finite-time future singularities to imagine different kinds of late time cosmological evolutions may happen in this model. \\
To start considering the finite-time future singularities in context of the massive gravity and massive bigravity, first we give short review of this issue in the GR case. Thus, by considering the standard FRW universe with the spatial curvature parameter ($\kappa$) and $8\pi G=c=1$, the Einstein field equations reduce to
\be
\label{Fbi16c}
3H^{2}=\rho-\frac{\kappa}{a^{2}},
\ee

\be
\label{Fbi17c}
\frac{\ddot{a}}{a}=-\left(\frac{\rho+3P}{6}\right),
\ee

Note that to solve the above equations we need the continuity equation, {\it i.e.} equation (\ref{Fbi7}).

So, to search the finite-time future cosmological singularities in the time interval $0<t<t_{s}$, we can write the following explicit expression for the scale factor which it has been studied in \cite{28}
\be
\label{Fbi16}
a(t)=1+\alpha t^{q} +\beta \left(t_{s}-t\right)^{n}.
\ee
Where $\alpha$, $\beta$, $q$ and $n$ are free positive constants which should be found. Also, the most general form of this solution in the vicinity of the singularity at $t=t_{s}$ for Friedmann equations has been considered in \cite{34}. The suggested expression for the scale factor (\ref{Fbi16}) fulfils our purposes about its behaviors in the cosmic time interval $0<t<t_{s}$ in which we expect the scale factor starts from zero then grows to a  $a_{s}$ value when $t\rightarrow t_{s}$.
By fixing the initial condition $a(0)=0$ at $t\rightarrow 0$, we get $\beta=-\frac{1}{t^{n}_{s}}$. Thus with the assumption $q$,~$n$$>0$ we obtain the following expression for the scale factor

\be
\label{Fbi17}
a(t)=1+\left(a_{s}-1\right)\left(\frac{t}{t_{s}}\right)^{q}-\left(1-\frac{t}{t_{s}}\right)^{n},
\ee
where $a_{s}\equiv a\left(t_{s}\right)$.
Then we find
\be
\label{Fbi18}
\dot{a}(t)=\frac{\left(-1+a_{s}\right)q\left(\frac{t}{t_{s}}\right)^{q}}{t}+\frac{n\left(1-\frac{t}{t_{s}}\right)^{-1+n}}{t_{s}}.
\ee

And so

\be
\label{Fbi19}
\ddot{a}(t)=\frac{\left(-1+a_{s}\right)\left(-1+q\right)q\left(\frac{t}{t_{s}}\right)^{q}}{t^{2}}-
\frac{\left(-1+n\right)n\left(1-\frac{t}{t_{s}}\right)^{-2+n}}{t^{2}_{s}}.
\ee

If $t\rightarrow t_{s}$ the above extracted equation reads
\be
\label{Fbi20c}
\ddot{a}\rightarrow q\left(q-1\right)\alpha t^{q-2}-\frac{n\left(n-1\right)}{t_{s}^{2}\left(1-\frac{t}{t_{s}}\right)^{2-n}}\rightarrow -\infty.
\ee
According to (\ref{Fbi17}), for $t\rightarrow t_{s}$ we should choose the intervals $1<n<2$ and $0<q<1$ we get $a\rightarrow a_{s}$, also $H_{s}$ and $\rho_{s}$ are finite but $P_{s}\rightarrow \infty$ because there is not any upper bound on the pressure profile. Moreover, for $2<n<3$ we have a finite $\ddot{a}$ but $\dddot{a}$ and also $P_{s}$ remains finite while $\dot{P_{s}}\rightarrow \infty$. In other words, there is an initial singularity when $t\rightarrow 0$ with $\rho\rightarrow\infty$ and $P\rightarrow \infty$. From (\ref{Fbi20c}) and (\ref{Fbi17c}) it can be seen that during the time interval $0<t<t_s$ the strong energy conditions hold ($\rho>0$ and $\rho +3P>0$). Therefor, this especial solution in an expanding universe indicates that it is contingent  to extend a ``sudden''  singularity at a finite-time future (type II), regardless of the sign of the spatial curvature parameter  \cite{34}.



Additionally, the ``big rip'' singularity (type I) can arise in GR and  modified gravity theories by the effect of matter fields which do not respect the dominant energy condition such as phantom or ghost field as $\rho+P<0$ \cite{27}.

it is worthwhile to note that, the extreme phantom equation of state ($\rho+P<0$ ) for generating a future singularity in a non-contracting universe is sufficient but not necessary, according to the first reference in \cite{28}. Therefore, this finite-time future singularity can appear in the expanding Friedmann universe without taking the condition $\rho+P<0$. Thus, the solution (\ref{Fbi16}) indicates that for an expanding universe, there is a possibility to develop a ``big-rip'' singularity (type I) in a finite-time in the future even the matter fields satisfy the strong energy conditions.\\
Other types of cosmological singularities (type III and IV) are different from the sudden finite-time future singularity in the sense that $\rho$ diverges. Usually, in these singularities the pressure profile and density have a non-realistic equation of state, {\rm e.g.} $P=-\rho-f(\rho)$ which they are studied in \cite{35}.
\\

In Table $1$ we summarize the types of possible future singularities in an expanding FRW universe considering the behavior of $a(t)$, $\rho(t)$ and $P(t)$ in a finite cosmological time \cite{35}. \\

 \vspace{5mm}
\begin{center}
{\scriptsize{ Table 1: }}\hspace{-2mm} {\scriptsize The possible finite-time future singularities for $t\rightarrow t_{s}$ }\\
    \begin{tabular}{|l| l |l| p{25mm}| }
    \hline
   {\footnotesize$~~~~~\textrm{Singularity Type~~~} $ }& {\footnotesize~~~ $a(t)$ }&{\footnotesize~~~~$\rho(t)~~~$ }&{\footnotesize~~~ $P(t)~$ } \\\hline
{\footnotesize ~~~~~~~~~~~~~$\textrm{\textbf{I}}$} & {~\footnotesize $a\rightarrow\infty$} & {\footnotesize~~ $\rho\rightarrow\infty$}&{\footnotesize~~~~~ $|P|\rightarrow\infty$ }\\\hline
 {\footnotesize ~~~~~~~~~~~~~$\textrm{\textbf{II}}$} & {~\footnotesize $a\rightarrow a_{s}$}& {\footnotesize  ~~~$\rho\rightarrow\rho_{s}$}&{\footnotesize~~~~~ $|P|\rightarrow\infty$ }
\\ \hline
{\footnotesize ~~~~~~~~~~~~$\textrm{\textbf{III}}$} & {~\footnotesize $a\rightarrow a_{s}$}& {\footnotesize  ~~~$\rho\rightarrow\infty$}&{\footnotesize~~~~~ $|P|\rightarrow\infty$ }\\\hline
{\footnotesize ~~~~~~~~~~~~$\textrm{\textbf{IV}}$} & {~\footnotesize $a\rightarrow a_{s}$}& {\footnotesize  ~~~$\rho\rightarrow0$}&{\footnotesize~~~~~ $|P|\rightarrow0$ } \\\hline
    \end{tabular}
\end{center}

According to table $1$, the type I singularity is known as ``big rip or cosmic doomsday'' explained in \cite{27} beside the type II, III and IV singularities are known as ``sudden'' \cite{28}, ``big freeze'' \cite{37} and ``big brake or big separation'' \cite{38} singularities, respectively. It is noticeable that in case IV singularity, however $a(t)$, $\rho(t)$ and $P(t)$ remain finite, the higher derivatives of $H(t)$ diverge.\\

So, to continue we try to classify the possible finite-time future singularities in the context of massive gravity theory with respect to $q$ and $n$ values as in the following classifications. It is worth noting, the spacial curvature $\kappa$ in this theory is free to be choused $\pm1$ or $0$ as studied in \cite{25} but according to those explained in the introduction we just work with $\kappa=-1$.\\

\textbf{Class I:}~~$\textbf{t}\rightarrow \textbf{t}_{s}$ \textbf{with} $\textbf{n}\in \left(\textbf{0}\textbf{,}\textbf{1}\right)$
\textbf{and} $\textbf{q}\in \left(\textbf{0}\textbf{,}\textbf{1}\right]$ \\

In this class we obtain

\be
\label{Fbi20}
a(t_{s})\rightarrow a_{s},~~~ \dot{a}(t_{s})\rightarrow +\infty,~~~H(t_{s})\rightarrow +\infty,~~~\ddot{a}(t_{s})\rightarrow +\infty,
\ee

\be
\label{Fbi21}
\rho (t_{s})\rightarrow \rho_{s}>0,~~~|P(t_{s})|\rightarrow \infty.
\ee

According to the table $1$, this is ``sudden singularity''. To more consider, we have plotted the evolution of the quantities $a(t)$, $\rho(t)$ and $P(t)$ in Fig. $1$, for two typical values of $q$. It can be seen that increasing $q$ value results in that $\rho(t)$ and $P(t)$ increase more rapidly for $t$ $\rightarrow$ $t_{s}$.\\\\

\begin{figure*}[ht]
  \centering
  \includegraphics[width=2in]{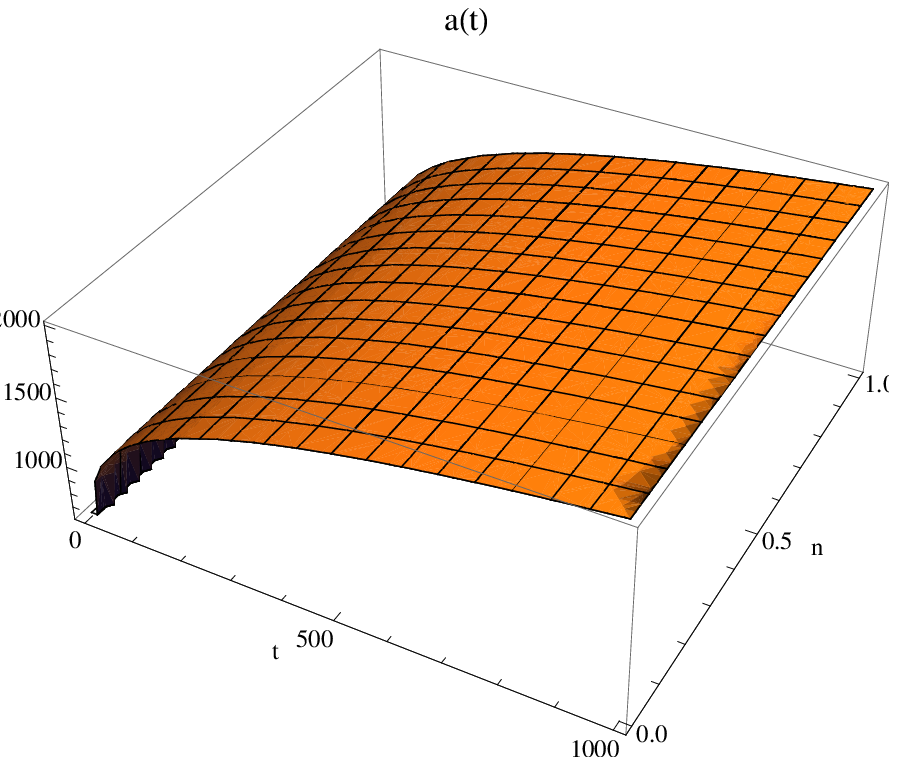}\hspace{1.9cm}
  \includegraphics[width=2in]{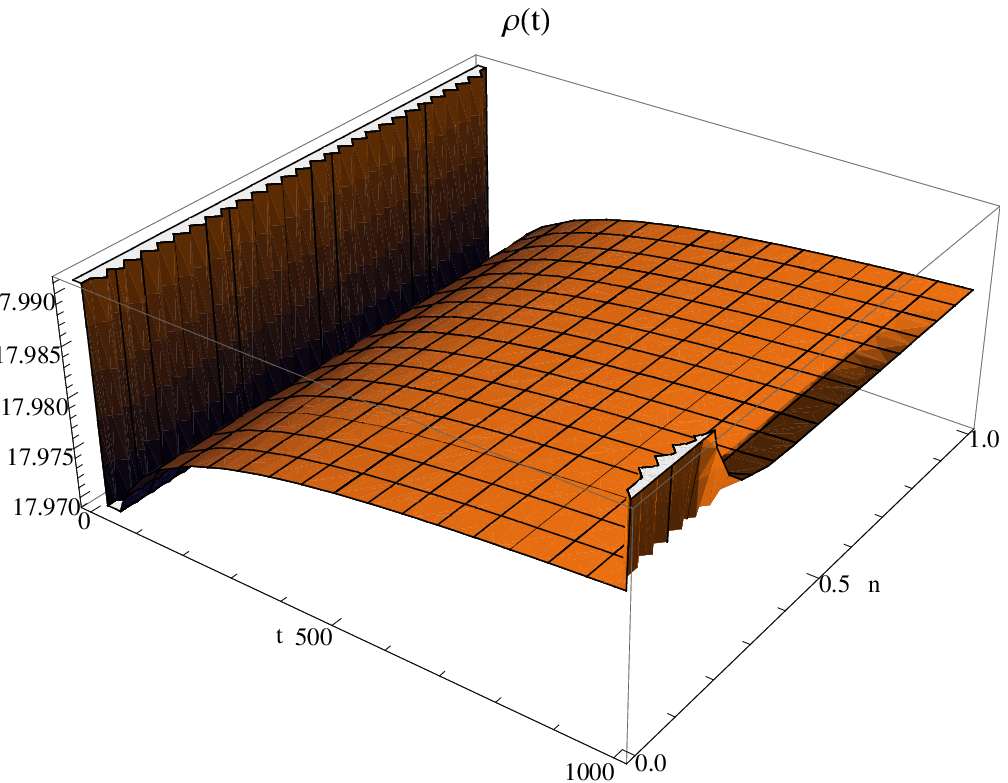}
  \includegraphics[width=2in]{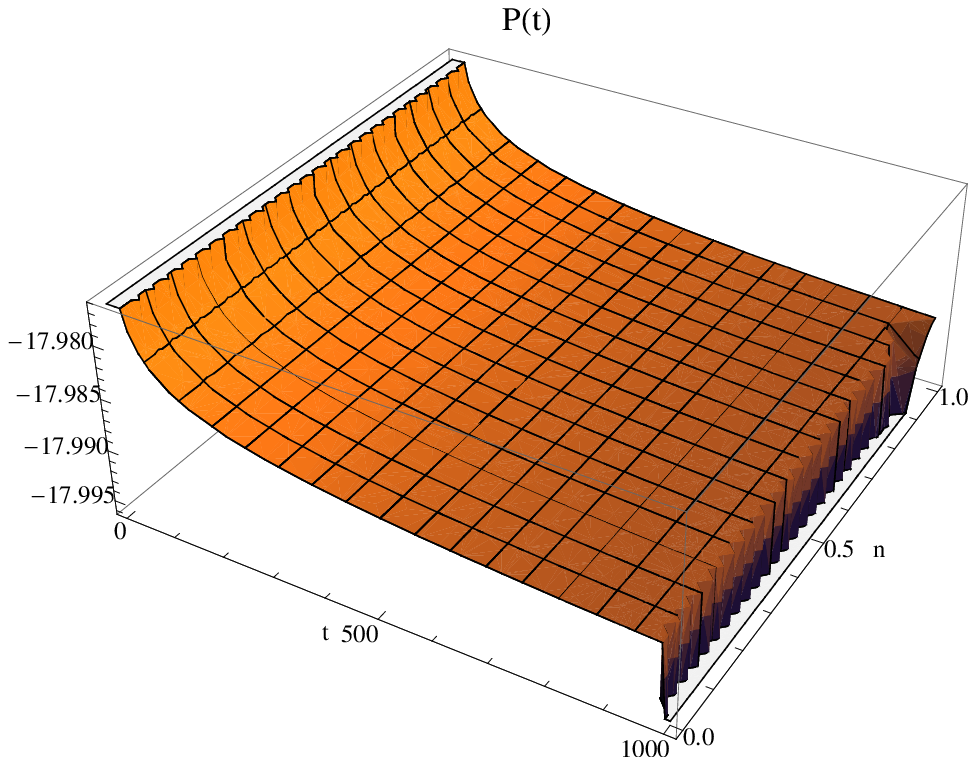}
  \includegraphics[width=2in]{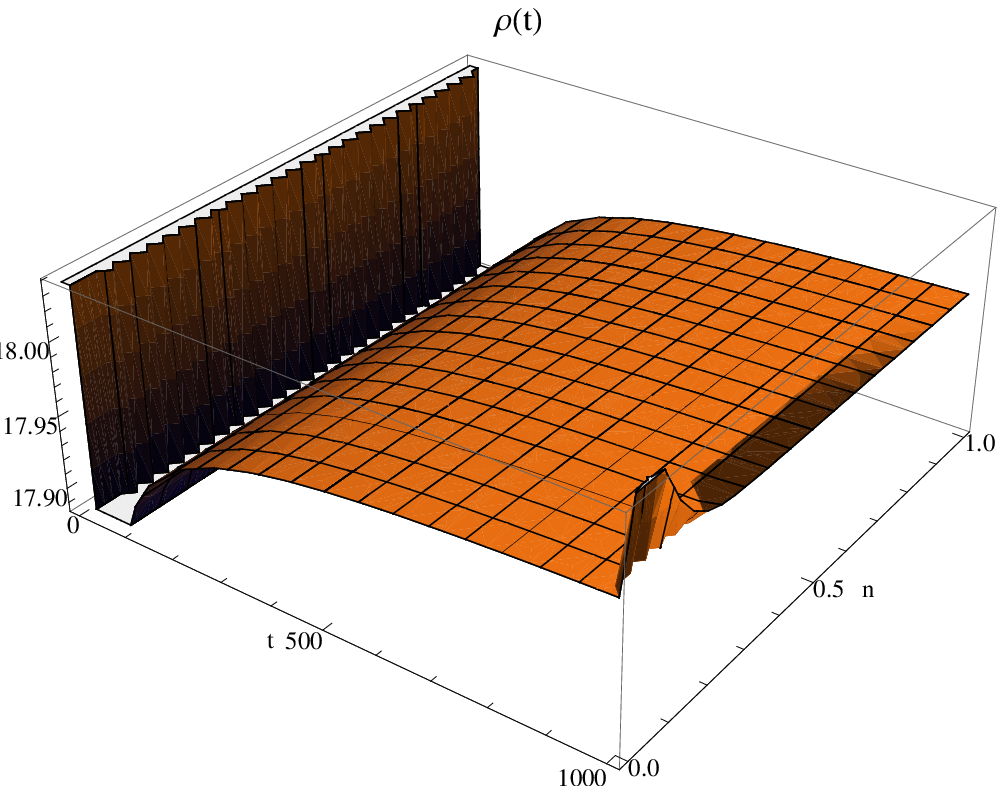}
  \includegraphics[width=2in]{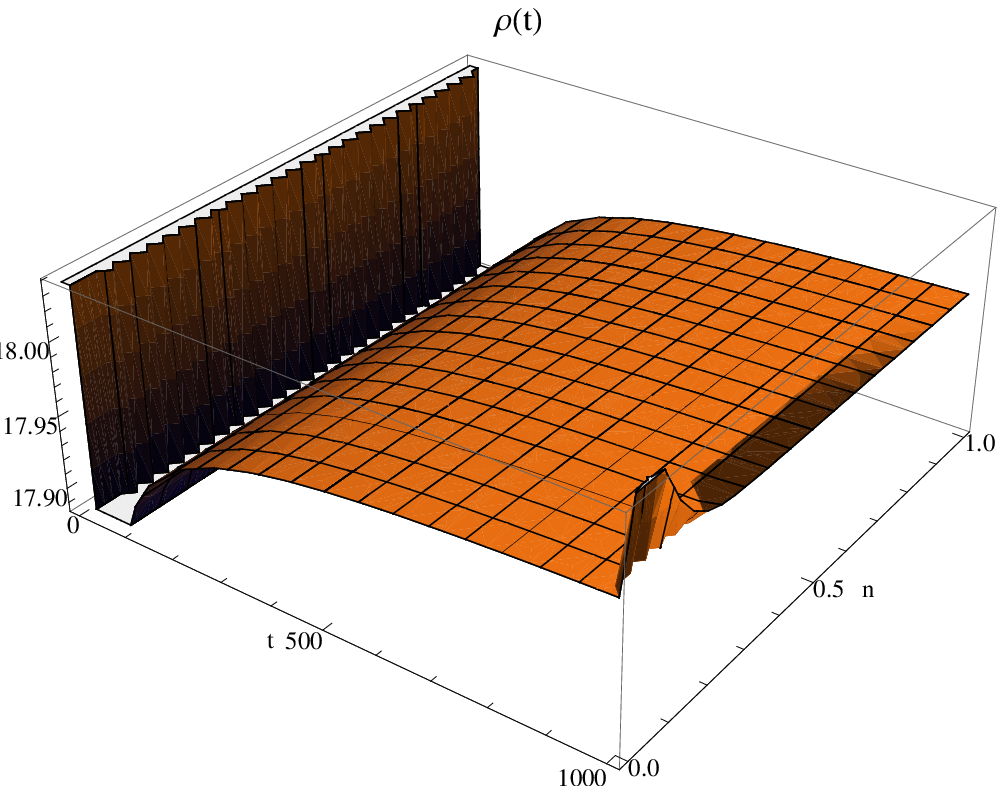}
  \includegraphics[width=2in]{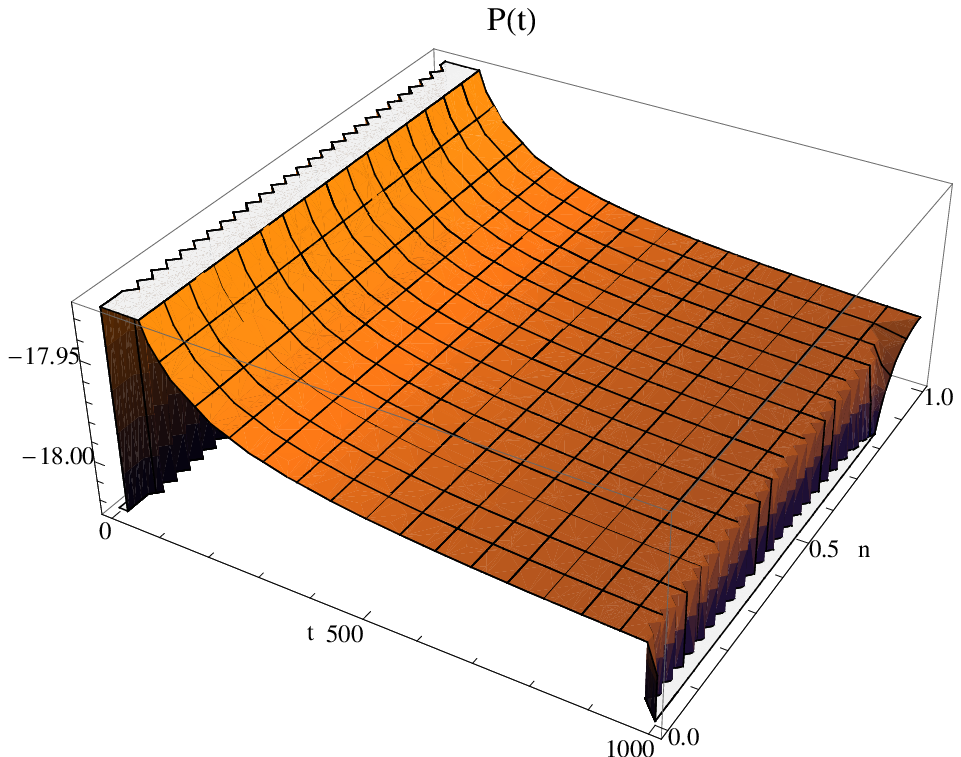}
    \caption{The scale factor $a(t)$, density $\rho(t)$ and pressure $P(t)$ evolution with two typical values  $q=0.2$ and $q=0.8$ in the upper and lower plots, respectively. We have fixed $a_{s}=2000$, $c_{4}=0.3$, $t_{s}=1000$, $M_{g}^{2}=1$ and $\kappa=-1$.}
  \label{stable}
\end{figure*}

\textbf{Class II:}~~$\textbf{t}\rightarrow \textbf{t}_{s}$ \textbf{with} $\textbf{n}=1$
\textbf{and} $\textbf{q}\in \left(\textbf{0}\textbf{,}\textbf{1}\right]$ \\\\\\\\
In this class one finds

\be
\label{Fbi21}
a(t_{s})\rightarrow a_{s},~~~ \dot{a}(t_{s})\rightarrow \dot{a}_{S}>0 ,~~~H(t_{s})\rightarrow H_{s}>0,~~~\ddot{a}(t_{s})\rightarrow \ddot{a}_{s}\leq 0,
\ee

\be
\label{Fbi22}
\rho (t_{s})\rightarrow \rho_{s}>0,~~~|P(t_{s})|\rightarrow P_{s}.
\ee

Clearly, there is no finite-time future singularity because all $a_{s}$, $\dot{a}_{s}$, $H_{s}$, $\ddot{a}_{s}$, $\rho_{s}$ and $P_{s}$ are finite. It is noticeable that for $q=1$ we obtain the zero acceleration $\ddot{a}_{s}=0$. Similar to class I we have plotted $a(t)$, $\rho(t)$ and $P(t)$ evolutions in Fig. $2$, for two typical values of $q$. We can see that by increasing the value of $q$ the slopes of the $\rho(t)$ and $P(t)$ for $t\rightarrow t_{s}$ is increasing.

\begin{figure*}[ht]
  \centering
  \includegraphics[width=2in]{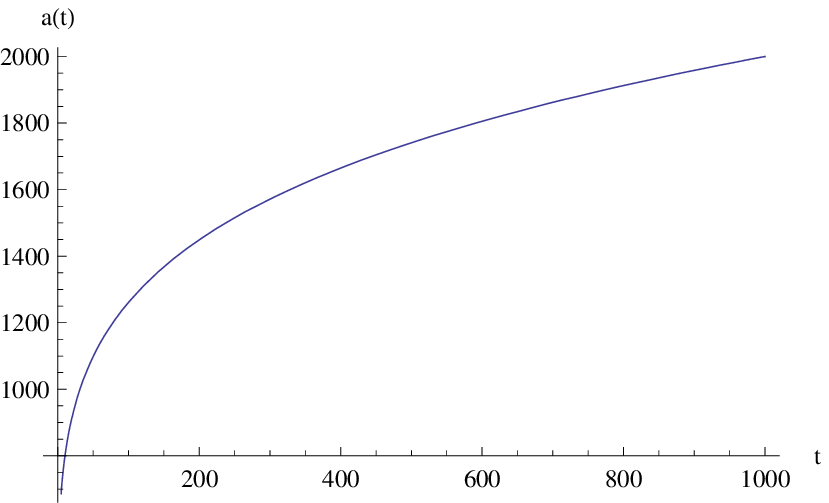}\hspace{1cm}
  \includegraphics[width=2in]{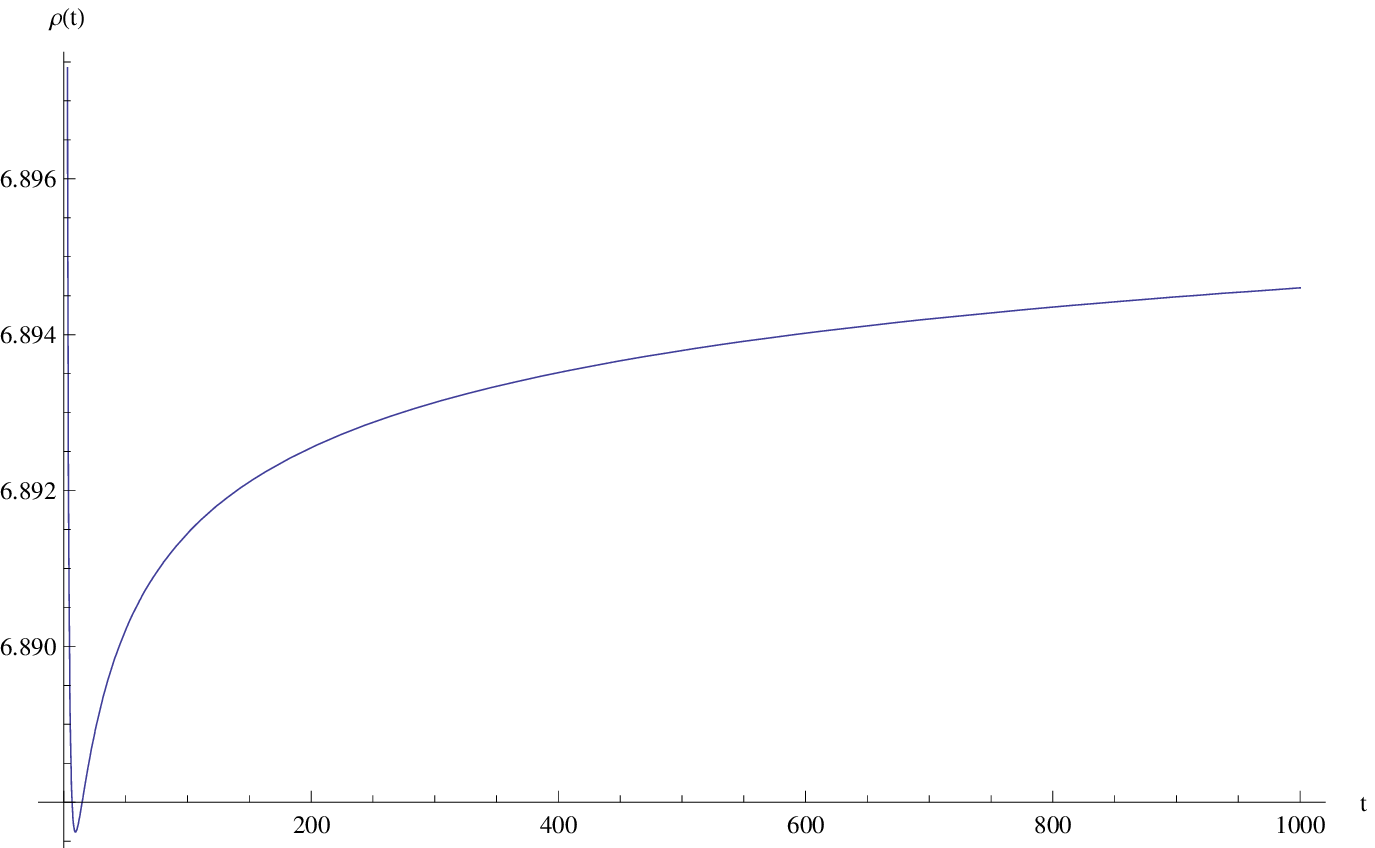}\hspace{1cm}
  \includegraphics[width=2in]{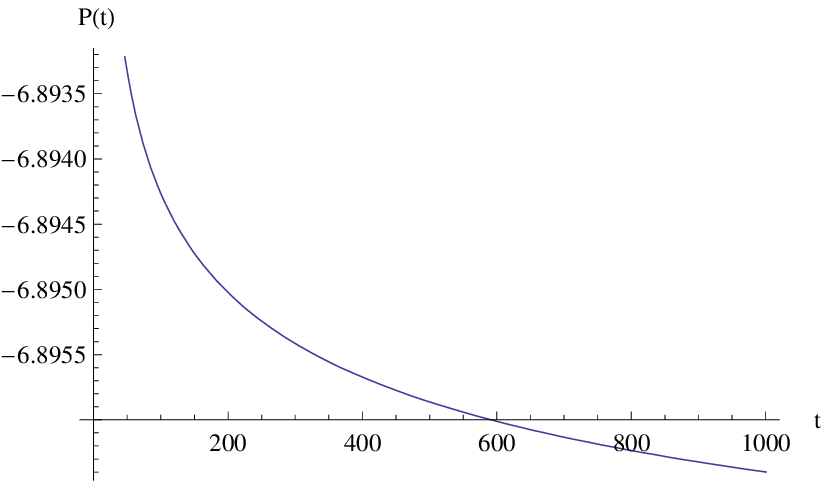}\hspace{0.1cm}
  \includegraphics[width=2in]{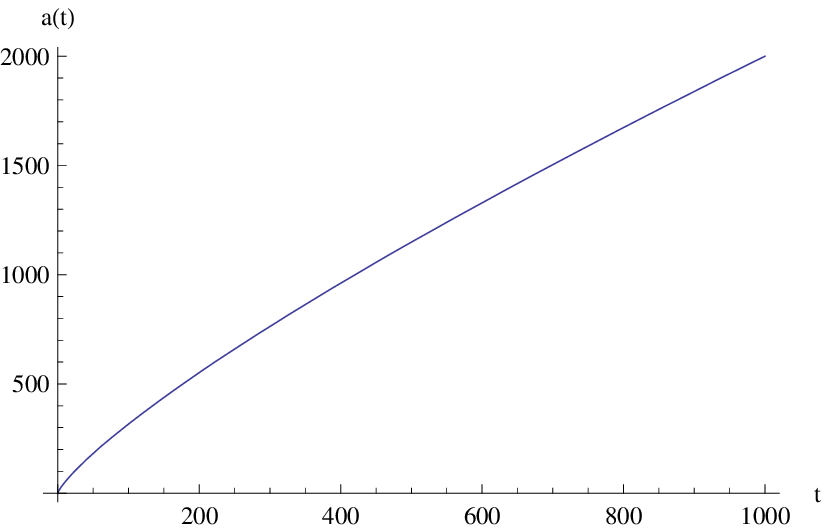}\hspace{0.7cm}
  \includegraphics[width=2in]{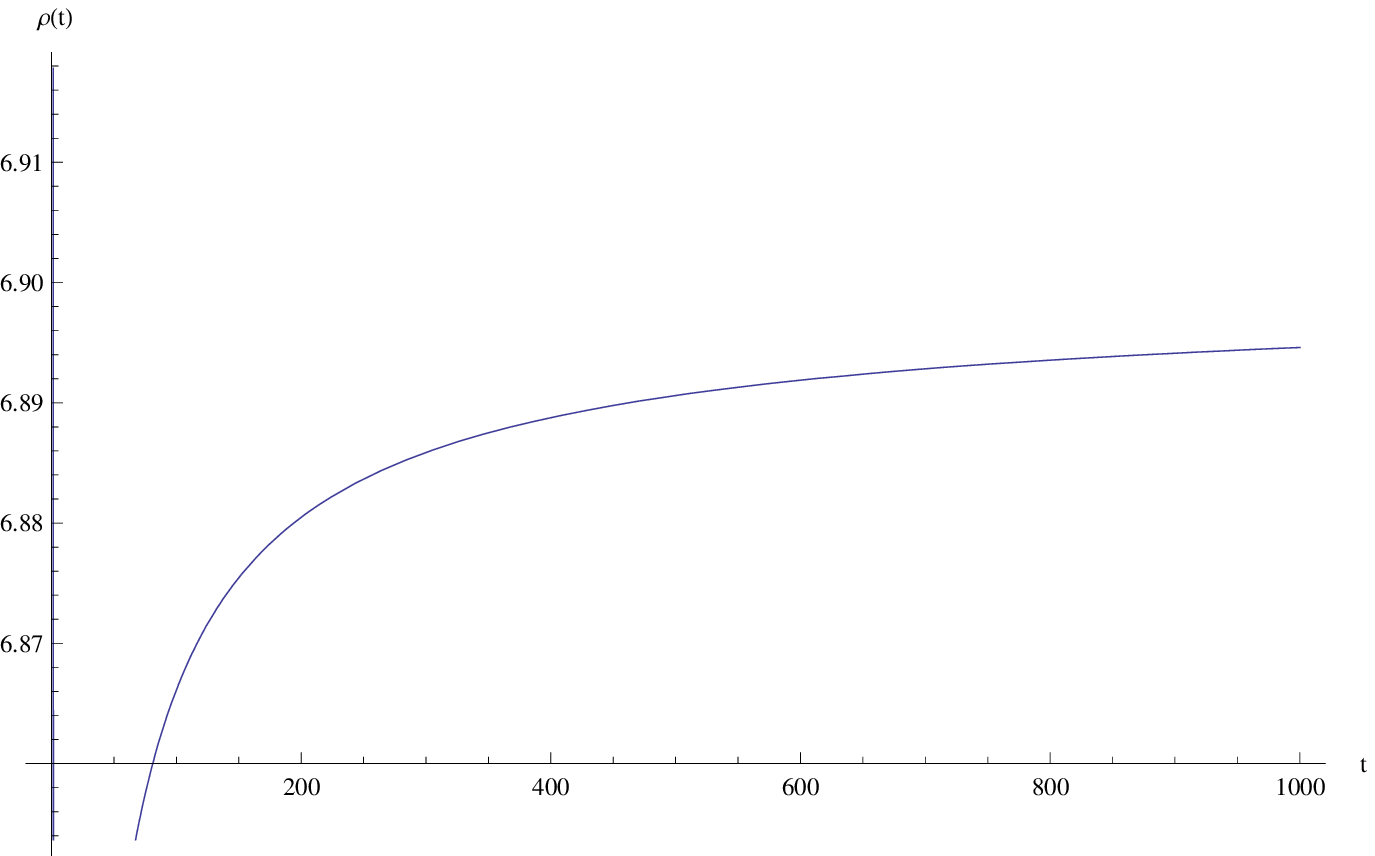}\hspace{0.7cm}
  \includegraphics[width=2in]{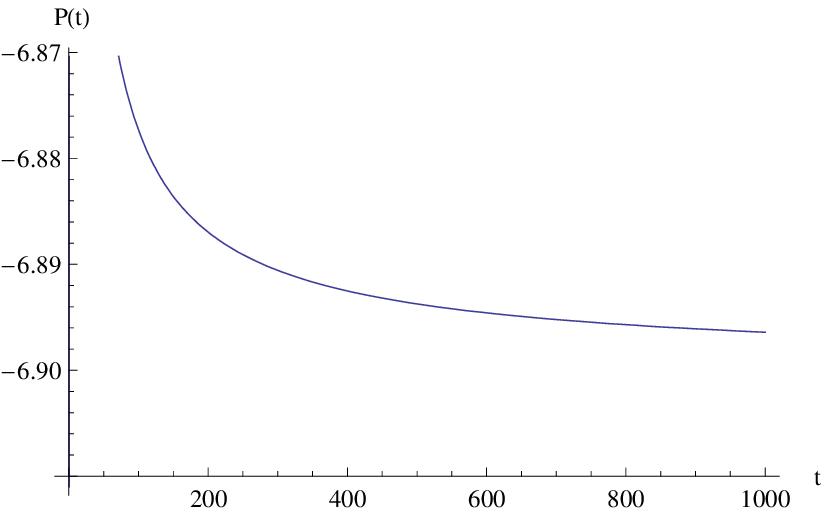}\hspace{0.7cm}
  \caption{The scale factor $a(t)$, density $\rho(t)$ and pressure $P(t)$ evolution by choosing $n=1$ with two typical values  $q=0.2$ and $q=0.8$ in the upper and lower plots, respectively. We take $a_{s}=2000$, $c_{4}=0.3$, $t_{s}=1000$, $M_{g}^{2}=1$ and $\kappa=-1$.}
  \label{stable}
\end{figure*}

\textbf{Class III:}~~$\textbf{t}\rightarrow \textbf{t}_{s}$ \textbf{with} $\textbf{n}\in \left(\textbf{1},\textbf{2}\right)$
\textbf{and} $\textbf{q}\in \left(\textbf{0}\textbf{,}\textbf{1}\right]$ \\\\\\\\
In this class one finds

\be
\label{Fbi23}
a(t_{s})\rightarrow a_{s},~~~ \dot{a}(t_{s})\rightarrow \dot{a}_{S}>0 ,~~~H(t_{s})\rightarrow H_{s}>0,~~~\ddot{a}(t_{s})\rightarrow -\infty,
\ee

\be
\label{Fbi24}
\rho (t_{s})\rightarrow \rho_{s}>0,~~~|P(t_{s})|\rightarrow P_{s}.
\ee

Apparently, there is no finite-time future singularity because $a_{s}$, $\rho_{s}$ and $P_{s}$ are finite. As a result, there is no finite-time future singularity in this class. Similar to previous cases, $a(t)$, $\rho(t)$ and $P(t)$ evolutions are plotted in Fig. $3$. By increasing $q$ the slopes of $\rho(t)$ and $P(t)$ is increasing for $t \rightarrow t_{s} $.

\begin{figure*}[ht]
  \centering
  \includegraphics[width=2in]{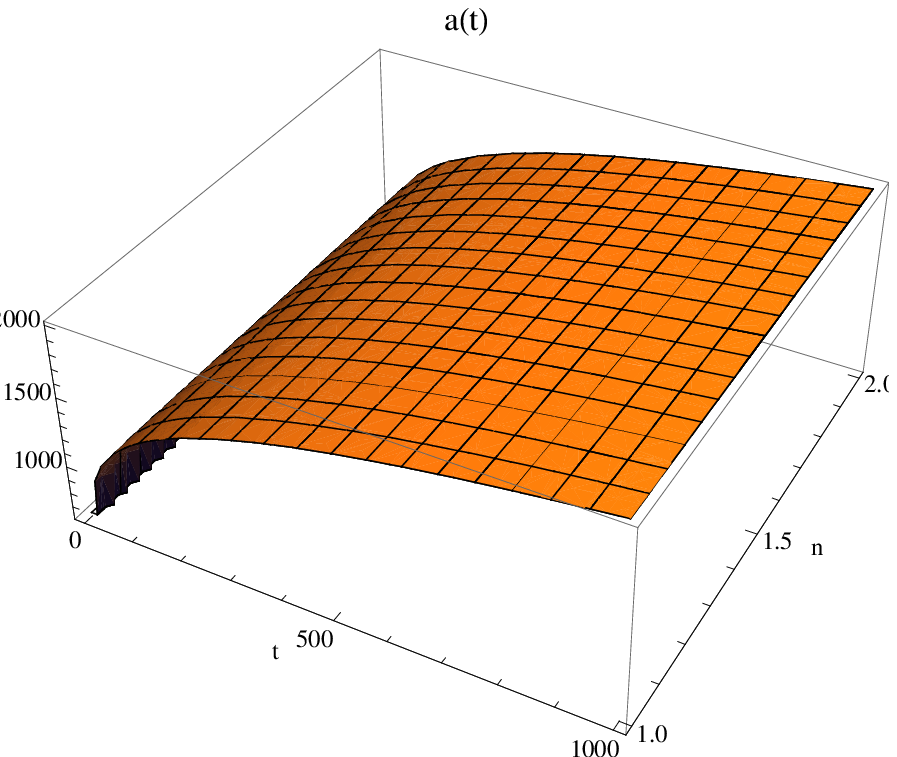}\hspace{1cm}
  \includegraphics[width=2in]{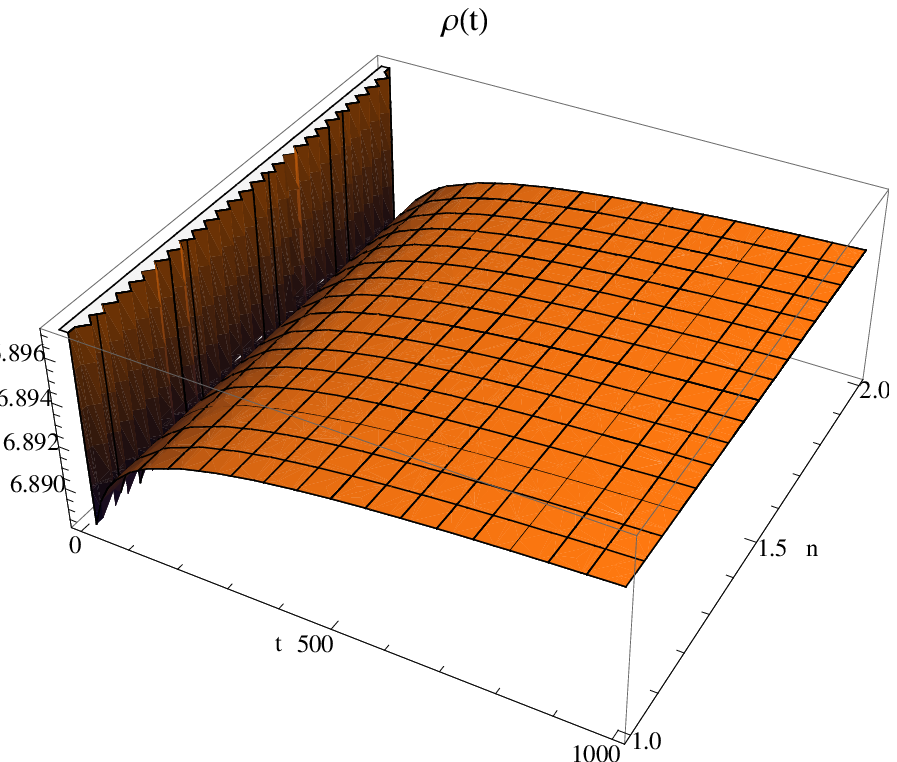}\hspace{1cm}
  \includegraphics[width=2in]{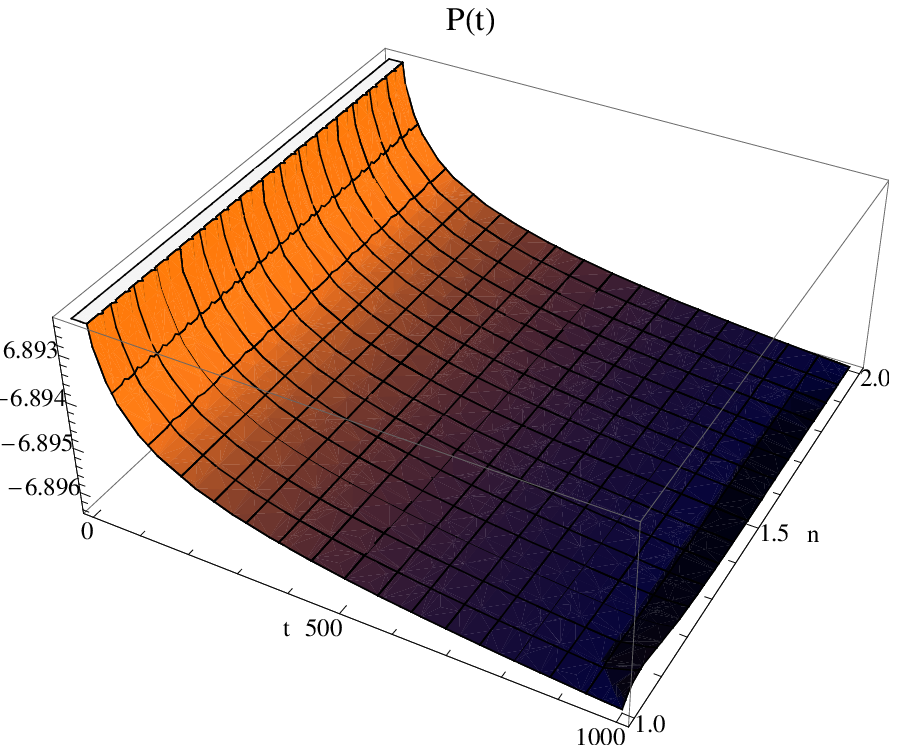}\hspace{0.1cm}
  \includegraphics[width=2in]{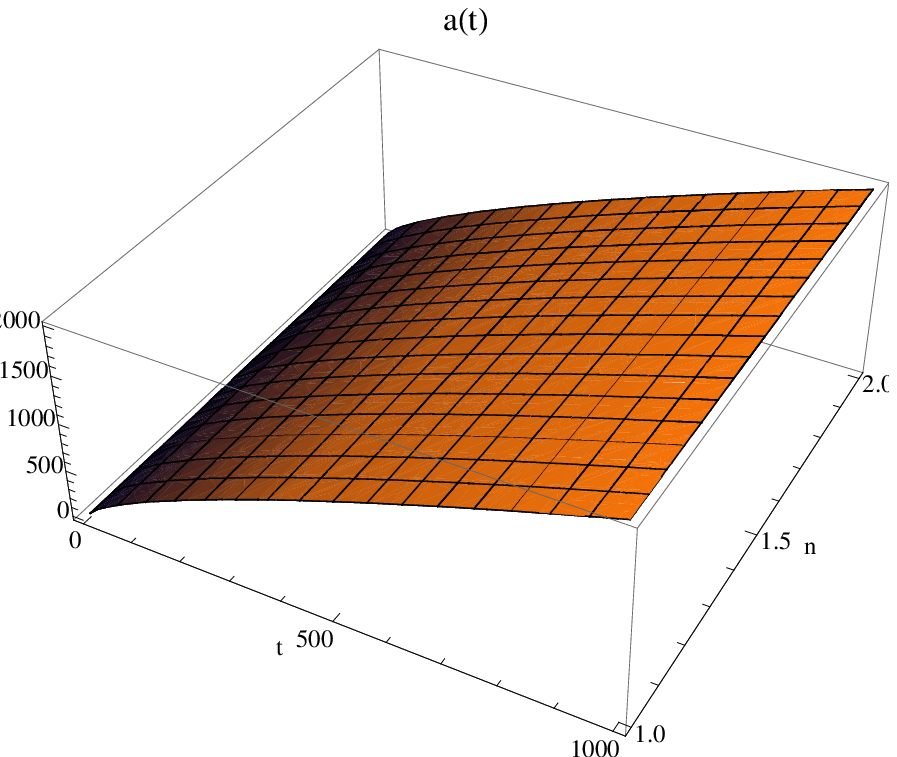}\hspace{0.7cm}
  \includegraphics[width=2in]{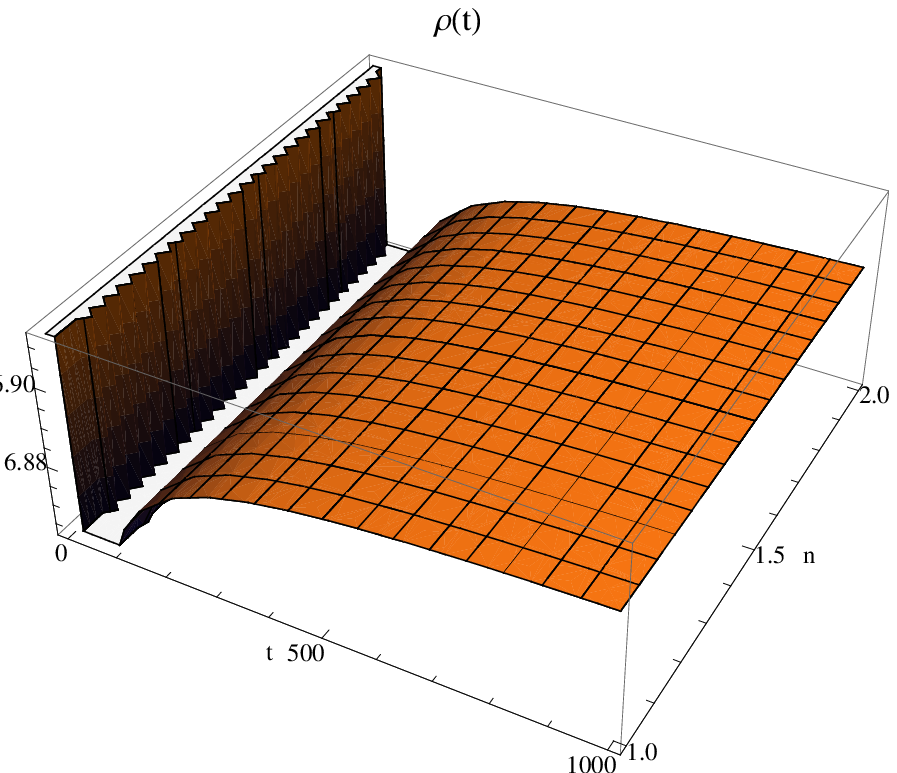}\hspace{0.7cm}
  \includegraphics[width=2in]{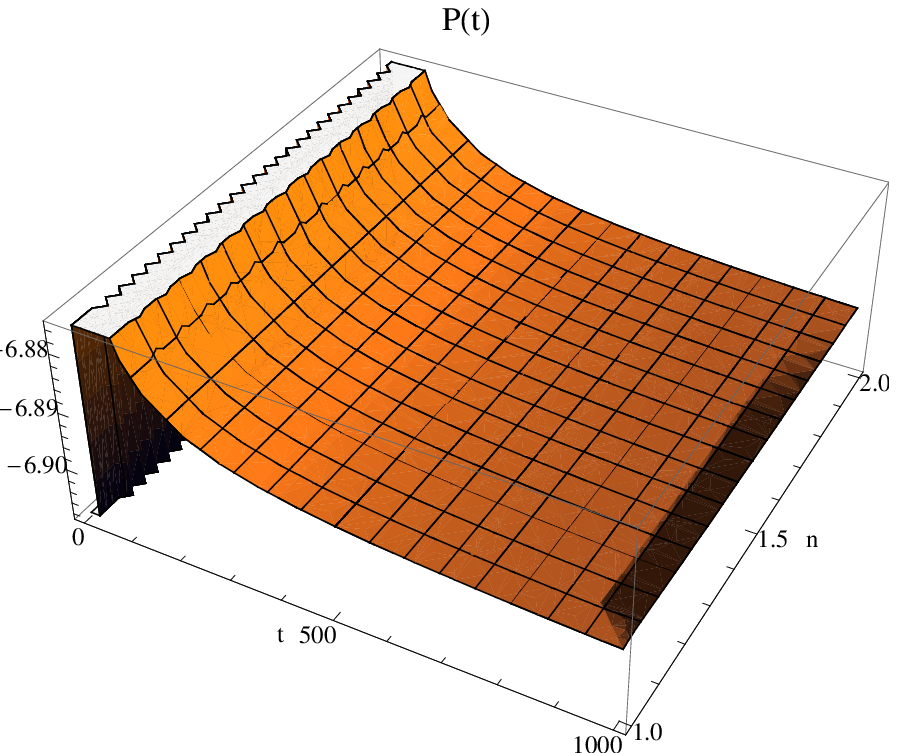}\hspace{0.7cm}
  \caption{The scale factor $a(t)$, density $\rho(t)$ and pressure $P(t)$ evolution with two typical values  $q=0.2$ and $q=0.8$ in the upper and lower plots, respectively. We have fixed $a_{s}=2000$, $c_{4}=0.3$, $t_{s}=1000$, $M_{g}^{2}=1$ and $\kappa=-1$.}
  \label{stable}
\end{figure*}

\textbf{Class IV:}~~$\textbf{t}\rightarrow \textbf{t}_{s}$ \textbf{with} $\textbf{n}\in \left[\textbf{2},\infty\right)$
\textbf{and} $\textbf{q}\in \left(\textbf{0}\textbf{,}\textbf{1}\right]$ \\\\\\
In this class we face with

\be
\label{Fbi25}
a(t_{s})\rightarrow a_{s},~~~ \dot{a}(t_{s})\rightarrow \dot{a}
_{S}>0 ,~~~H(t_{s})\rightarrow H_{s}>0,~~~\ddot{a}(t_{s})\rightarrow \ddot{a}_{s}\leq 0 ,
\ee

\be
\label{Fbi26}
\rho (t_{s})\rightarrow \rho_{s}>0,~~~|P(t_{s})|\rightarrow P_{s}.
\ee

Obviously, in this class there is not any singularity and all quantities are finite. The behavior of the scale factor $a(t)$, density $\rho(t)$ and pressure $P(t)$ for two typical values of $q$ are plotted in Fig. $4$.\\

\begin{figure*}[ht]
  \centering
  \includegraphics[width=2in]{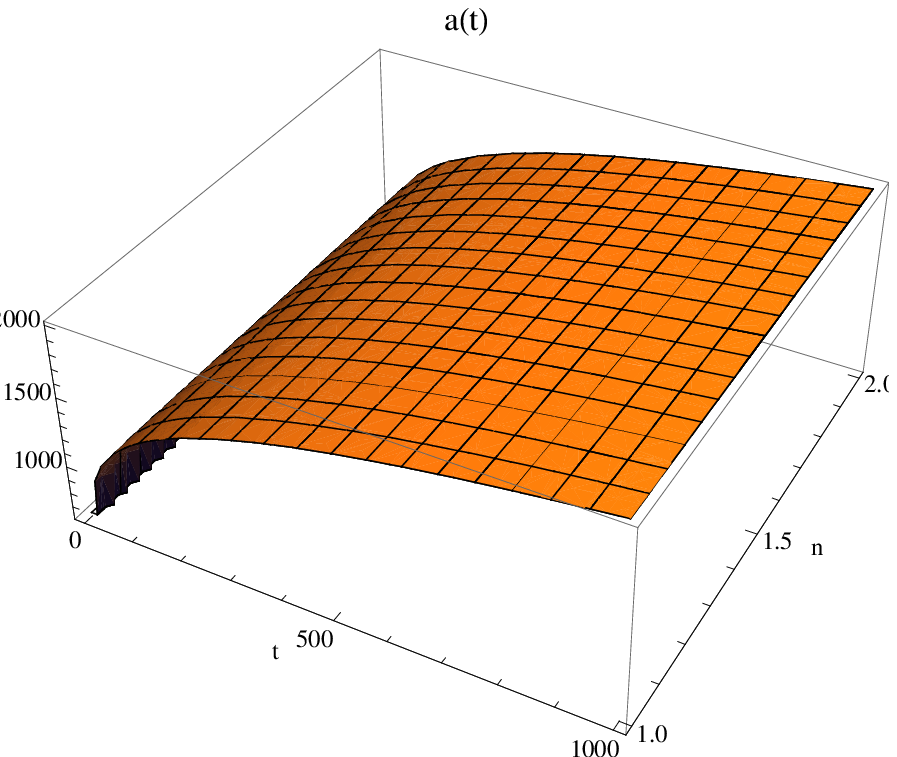}\hspace{1cm}
  \includegraphics[width=2in]{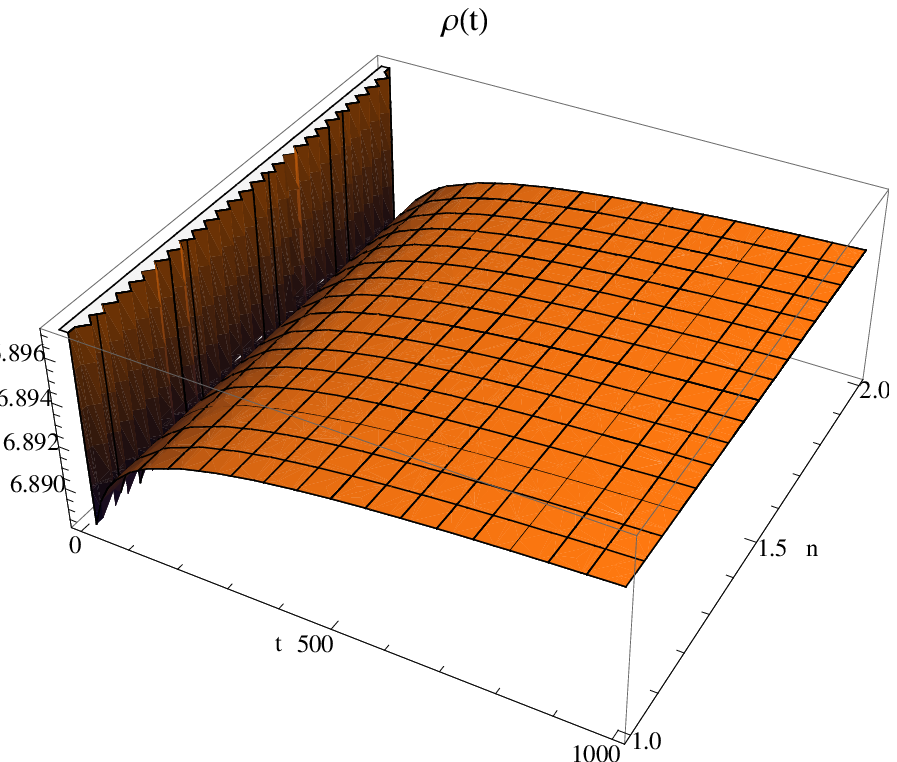}\hspace{1cm}
  \includegraphics[width=2in]{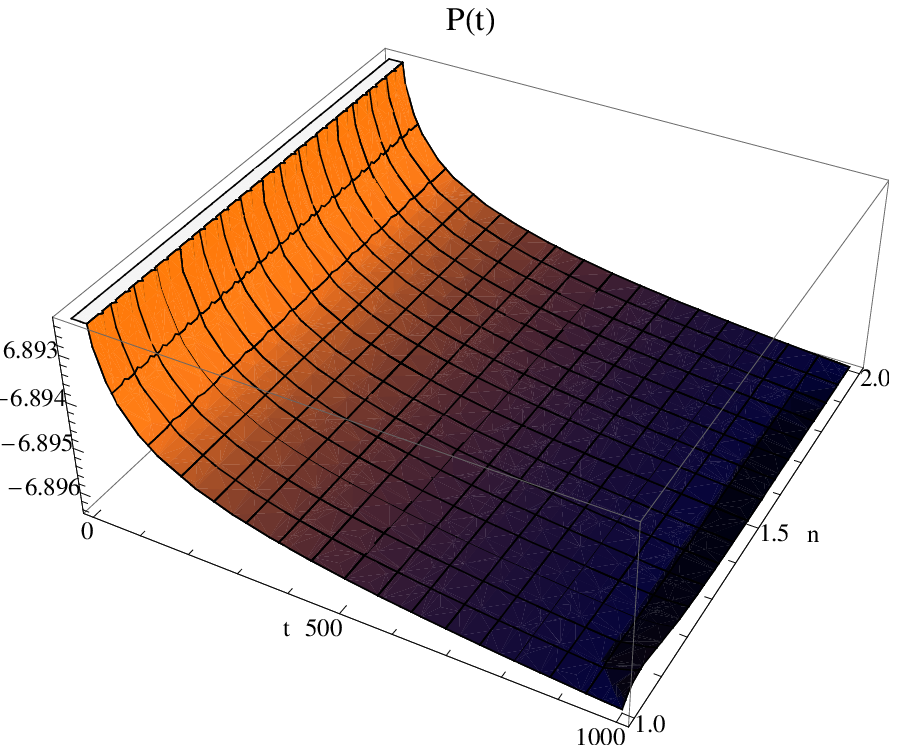}\hspace{0.1cm}
  \includegraphics[width=2in]{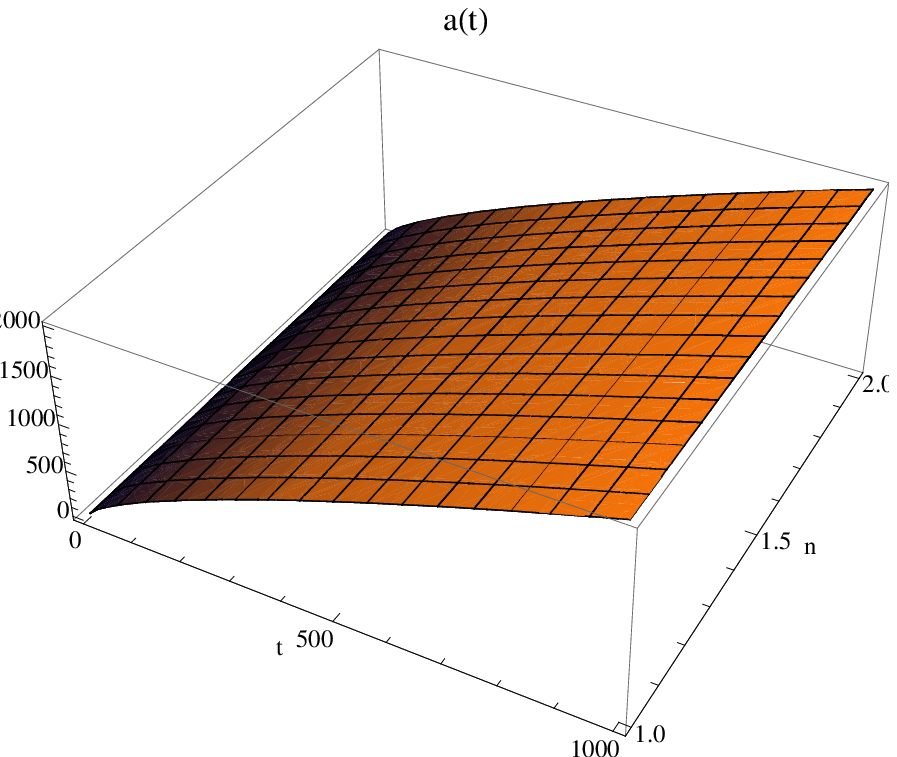}\hspace{0.7cm}
  \includegraphics[width=2in]{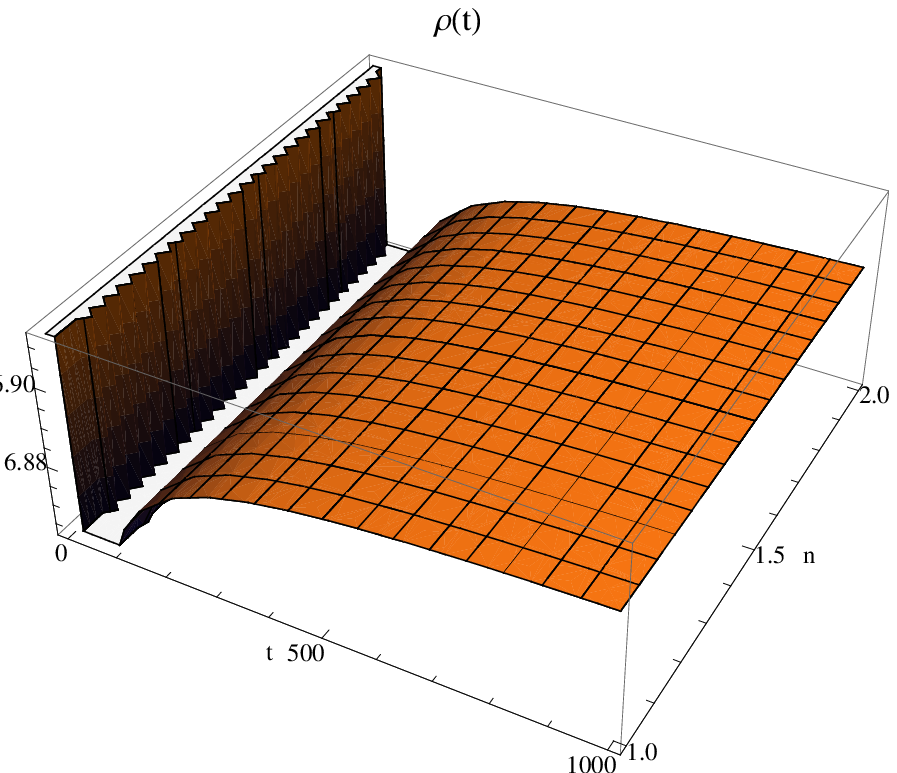}\hspace{0.7cm}
  \includegraphics[width=2in]{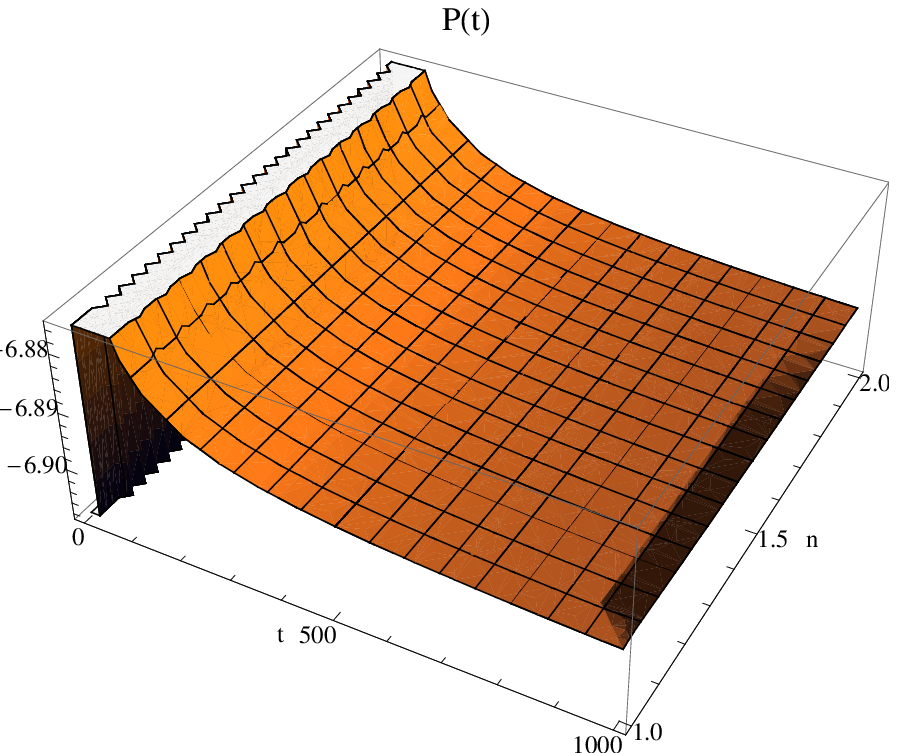}\hspace{0.7cm}
  \caption{The scale factor $a(t)$, density $\rho(t)$ and pressure $P(t)$ evolution with two typical values  $q=0.2$ and $q=0.8$ in the upper and lower plots, respectively. We have fixed $a_{s}=2000$, $c_{4}=0.3$, $t_{s}=1000$, $M_{g}^{2}=1$ and $\kappa=-1$.}
  \label{stable}
\end{figure*}

In the next section, we introduce the massive bigravity theory by extracting the field equations. Moreover we find the field equations of the most simplified version of this theory called minimal massive bigravity model to consider the finite-time future singularities in the cosmological backgrounds. We show that this simplified massive bigravity model suffers from ``sudden'' and ``big brake'' singularities.\\

\section{Friedmann equations in the Massive Bigravity theory \label{Sec4}}

In this section we review the ghost free massive bigravity model and its equations of motion in the background of FRW universe. The most general action for two interacting metrics $g_{\mu\nu}$ and $f_{\mu\nu}$ with a non derivative potential determined by the requirement of the absence of the Boulware-Deser ghost considering the coupling to matter via the matter lagrangian $\mathcal{L}_{m}$ \cite{16}, is

\begin{align}\label{Fbi27}
S=-\frac{M_{g}^{2}}{2}\int dx^{4}\sqrt{-det g} R(g)-\frac{M_{f}^{2}}{2}\int dx^{4}\sqrt{-det f} R(f)+m^{2}M_{g}^{2}\int dx^{4}\sqrt{-det g}\sum^{4}_{n=0} \beta_{n}e_{n}\left(g^{-1}f\right)+\int dx^{4}\sqrt{-det g}\mathcal{L}_{m},
\end{align}

that $\beta_{n}$ are free parameters and $e_{n}(X)$'s  are elementary symmetric polynomials of the eigenvalues of $X$:

\begin{align}\label{Fbi28}
e_{0}(X)=&1,~~e_{1}(X)=[X],~~e_{2}(X)=\frac{1}{2}\left([X]^{2}-[X^{2}]\right),\nn
e_{3}(X)=&\frac{1}{6}\left([X]^{3}-3[X][X^{2}]+2[X^{3}]\right),\nn
e_{4}(X)=&\frac{1}{24}\left([X]^{4}-6[X]^{2}[X^{2}]+3[X^{2}]^{2}+8[X][X^{3}]-6[X^{4}]\right),\nn
e_{i}(X)=&0~~{\rm for}~~i>4.
\end{align}
The square root matrix $\sqrt{g^{-1}f}$ is defined by $\left(\sqrt{g^{-1}f}\right)^{\mu}~_{\rho}\left(\sqrt{g^{-1}f}\right)^{\rho}~_{\nu}=g^{\mu\rho}f_{\rho\nu}=X^{\mu}~_{\nu}$.

We consider the FRW metric with spatial curvature $\kappa$, for both line elements as follows
\be
\label{Fbi29}
ds_{g}^{2}=-dt^{2}+a(t)^{2}\left(\frac{dr^{2}}{1-\kappa r^{2}}+r^{2}d\theta^{2}+r^{2}\sin^{2}\theta d\varphi^{2}\right),
\ee
\be
\label{Fbi30}
ds_{f}^{2}=-X(t)^{2}dt^{2}+Y(t)^{2}\left(\frac{dr^{2}}{1-\kappa r^{2}}+r^{2}d\theta^{2}+r^{2}\sin^{2}\theta d\varphi^{2}\right).
\ee

Where $a(t)$ is the cosmic scale factor related to $g_{\mu\nu}$ and $Y(t)$ is the one related to $f_{\mu\nu}$. Obviously, $X(t)$ (the lapse function of $f_{\mu\nu}$ metric) is a function of time and note that we are not allowed to choose $X=1$ nor $X=Y$.
According to \cite{19} in which, the Bianchi constraint (implied by general covariance) has been extracted in details,  turns out

\be
\label{Fbi31}
X=\frac{\dot{Y}}{\dot{a}}=\frac{dY}{da},
\ee

which leads to leaving just two free functions to work with. Using this relation beside the ansatz (\ref{Fbi29}) and (\ref{Fbi30}) give us the modified Friedmann equations of the $g_{\mu\nu}$ metric

\be
\label{Fbi32}
\rho=M_{g}^{2}\left(3\left(\frac{\dot{a}}{a}\right)^{2}+\frac{3\kappa}{a^{2}}-m^{2}\left(\beta_{0}+3\beta_{1}\frac{Y}{a}+3\beta_{2}\frac{Y^{2}}{a^{2}}+
\beta_{3}\frac{Y^{3}}{a^{3}}\right)\right),
\ee
\be
\label{Fbi33}
P=M_{g}^{2}\left(-2\frac{\ddot{a}}{a}-\frac{\dot{a}^{2}}{a^{2}}-\frac{\kappa}{a^{2}}+m^{2}\left(\beta_{0}+2\beta_{1}\left(\frac{Y}{a}+
X\right)+\beta_{2}\left(\frac{Y^{2}}{a^{2}}+\frac{2YX}{a}\right)+
\beta_{3}\frac{Y^{2}X}{a^{2}}\right)\right).
\ee

It should be noted that here we have the source structure $T^{0}_{0}=-\rho$, $T^{1}_{1}=P$ because we have assumed a perfect fluid source $T_{\mu\nu}=\left(\rho+P\right)u_{\mu}u_{\nu}+Pg_{\mu\nu}$ with $T^{1}_{1}=T^{2}_{2}=T^{3}_{3}$ (consistent with the symmetries of the space time). As $m^{2}\rightarrow 0$, these $g_{\mu\nu}$ Friedmann equations reduce to the ordinary Friedmann equations of cosmology but the cosmological solutions will not always be well defined in this limit. Moreover, we can find the modified Friedmann equations of the metric $f_{\mu\nu}$ as follows

\be
\label{Fbi34}
-3\left(\frac{\dot{a}}{Y}\right)^{2}-\frac{3\kappa}{Y^{2}}+\frac{m^{2}}{M_{*}^{2}}\left(\beta_{4}+3\beta_{3}\frac{a}{Y}+
3\beta_{2}\frac{a^{2}}{Y^{2}}+\beta_{1}\frac{a^{3}}{Y^{3}}\right)=0,
\ee

\be
\label{Fbi35}
-2\frac{\ddot{a}}{YX}-\left(\frac{\dot{a}}{Y}\right)^{2}-\frac{\kappa}{Y^{2}}+\frac{m^{2}}{M_{*}^{2}}\left(
\beta_{4}+\beta_{3}\left(2\frac{a}{Y}+\frac{1}{X}\right)+\beta_{2}\left(\frac{a^{2}}{Y^{2}}+
\frac{2a}{YX}\right)+\beta_{1}\frac{a^{2}}{Y^{2}X}\right)=0.
\ee

In which we have the dimensionless ratio of Planck masses as

\be
\label{Fbi35f}
M_{*}^{2}\equiv \frac{M_{f}^{2}}{M_{g}^{2}}.
\ee

Meanwhile, for simplicity we assume

\be
\label{Fbi36}
\gamma\equiv\frac{Y}{a}.
\ee

Therefore, equations (\ref{Fbi32}) and (\ref{Fbi34}) can be read as

\be
\label{Fbi37}
\frac{\beta_{3}}{3}\gamma^{3}+\beta_{2}\gamma^{2}+\beta_{1}\gamma+\frac{\beta_{0}}{3}+\frac{\rho}{3m^{2}M_{g}^{2}}-\frac{H^{2}}{m^{2}}-
\frac{\kappa}{m^{2}a^{2}}=0,
\ee
and
\be
\label{Fbi38}
\frac{\beta_{4}}{3M_{*}^{2}}\gamma^{2}+\frac{\beta_{3}}{M_{*}^{2}}\gamma+\frac{\beta_{1}}{3M_{*}^{2}}\frac{1}{\gamma}+\frac{\beta_{2}}{M_{*}^{2}}-
\frac{H^{2}}{m^{2}}-\frac{\kappa}{m^{2}a^{2}}=0.
\ee

Obviously, by using two above equations then eliminating $H^{2}$ we find the following quadratic equation for $\gamma$

\be
\label{Fbi39}
\frac{\beta_{3}}{3}\gamma^{4}+\left(\beta_{2}-\frac{\beta_{4}}{3M_{*}^{2}}\right)\gamma^{3}+\left(\beta_{1}-\frac{\beta_{3}}{M_{*}^{2}}\right)
\gamma^{2}+\left(\frac{\rho}{3m^{2}M_{g}^{2}}+\frac{\beta_{0}}{3}-\frac{\beta_{2}}{M_{*}^{2}}\right)\gamma-\frac{\beta_{1}}{3M_{*}^{2}}=0.
\ee

\subsection{The minimal model with $\beta_{1}=\beta_{3}=0$ \label{Sec3}}

Although, setting only $\beta_{3}=0$, in the action (\ref{Fbi27}) eliminates the highest order interaction term in $\sqrt{g^{-1}f}$, because of the invariance of this action under the following exchange

\be
\label{Fbi40}
g\leftrightarrow f,~~~~~~~\beta_{n} \leftrightarrow \beta_{4-n},~~~~~~~~~M_{g}\leftrightarrow M_{f},~~~~~~~~~m^{2}\leftrightarrow \frac{m^{2}M_{g}^{2}}{M_{f}^{2}},
\ee

we still have a cubic order interaction term in $\sqrt{f^{-1}g}$ which we eliminate it by choosing $\beta_{1}=0$ to obtain the minimal massive bigravity model which is the simplest in the class. Actually, in this class we have $\beta_{2}<\frac{\beta_{4}}{3M_{*}^{2}}$. As a result, for these values equation (\ref{Fbi39}) gives

\be
\label{Fbi41}
\gamma^{2}=\frac{\frac{\rho}{m^{2}M_{g}^{2}}+\beta_{0}-3\beta_{2}M_{*}^{-2}}{\beta_{4}M_{*}^{-2}-3\beta_{2}}.
\ee

Referring to equations (\ref{Fbi34}) and (\ref{Fbi35}) with $\beta_{1}=\beta_{3}=0$, we can find the solution

\be
\label{Fbi42}
\frac{\dot{Y}}{\dot{a}}=X(t)=\frac{\frac{-2a\ddot{a}}{\gamma}+\frac{2m^{2}a^{2}\beta_{2}}{\gamma M_{*}^{2}}}{\frac{\dot{a}^{2}}{\gamma^{2}}+\frac{\kappa}{\gamma^{2}}-
\frac{m^{2}a^{2}\beta_{4}}{M_{*}^{2}}-\frac{m^{2}a^{2}\beta_{2}}{M_{*}^{2}\gamma^{2}}}.
\ee

This relation helps us to simplify the pressure equation (\ref{Fbi33}) to become able to plot the functions of $P(t)$ and $\rho(t)$ to analyze their behaviors. \\

\section{Future cosmological singularities in Massive Bigravity Theory\label{Sec3}}

As section III, we are going to follow that approach again to study the possible finite-time future singularities in minimal massive bigravity universe. Considering (\ref{Fbi17}), (\ref{Fbi18}) and (\ref{Fbi19}) we calculate some classifications in the following.\\

\textbf{Class I:}~~$\textbf{t}\rightarrow \textbf{t}_{s}$ \textbf{with} $\textbf{n}\in \left(\textbf{0}\textbf{,}\textbf{1}\right)$
\textbf{and} $\textbf{q}\in \left(\textbf{0}\textbf{,}\textbf{1}\right]$ \\

In this class we can write

\be
\label{Fbi43}
a(t_{s})\rightarrow a_{s},~~~ \dot{a}(t_{s})\rightarrow \dot{a}_{s}>0,~~~H(t_{s})\rightarrow H_{s} >0,~~~\ddot{a}(t_{s})\rightarrow \ddot{a}_{s}\geq 0,
\ee

\be
\label{Fbi44}
\rho (t_{s})\rightarrow 0,~~~|P(t_{s})|\rightarrow 0.
\ee

This result represents the case IV singularity as it mentioned in Table $1$, namely ``big brake or big separation''. We have plotted the evolution of the quantities $a(t)$, $\rho(t)$ and $P(t)$ in Fig. $5$, for two typical values of $q$. We can see that increasing $q$ value results in that $\rho(t)$ and $P(t)$ increase more rapidly for $t$ $\rightarrow$ $t_{s}$.\\\\

\begin{figure*}[ht]
  \centering
  \includegraphics[width=2in]{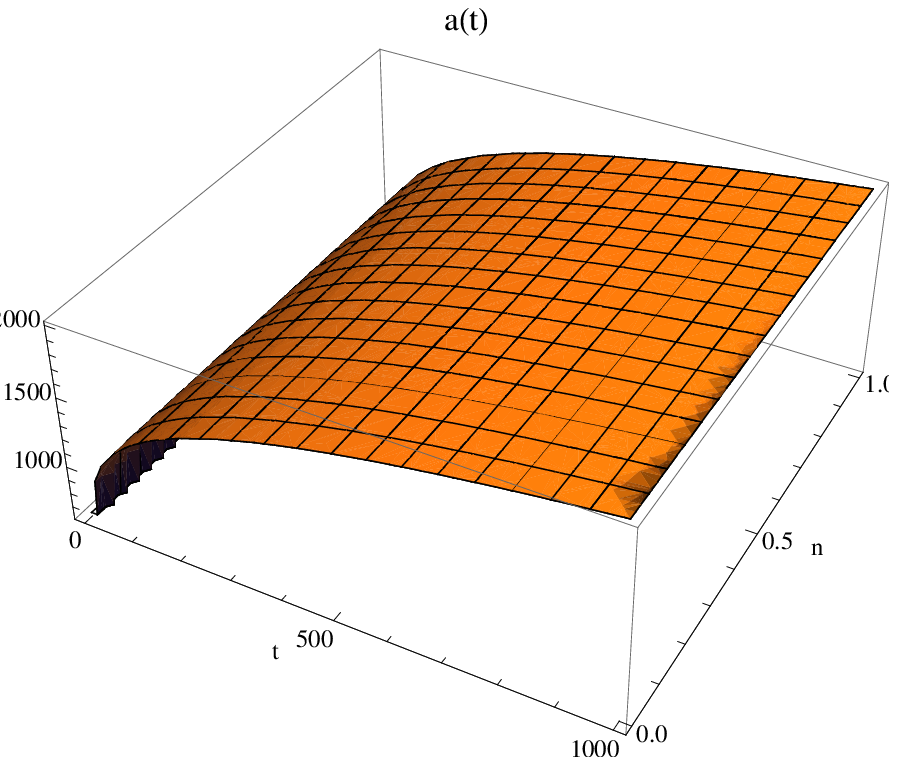}\hspace{1cm}
  \includegraphics[width=2in]{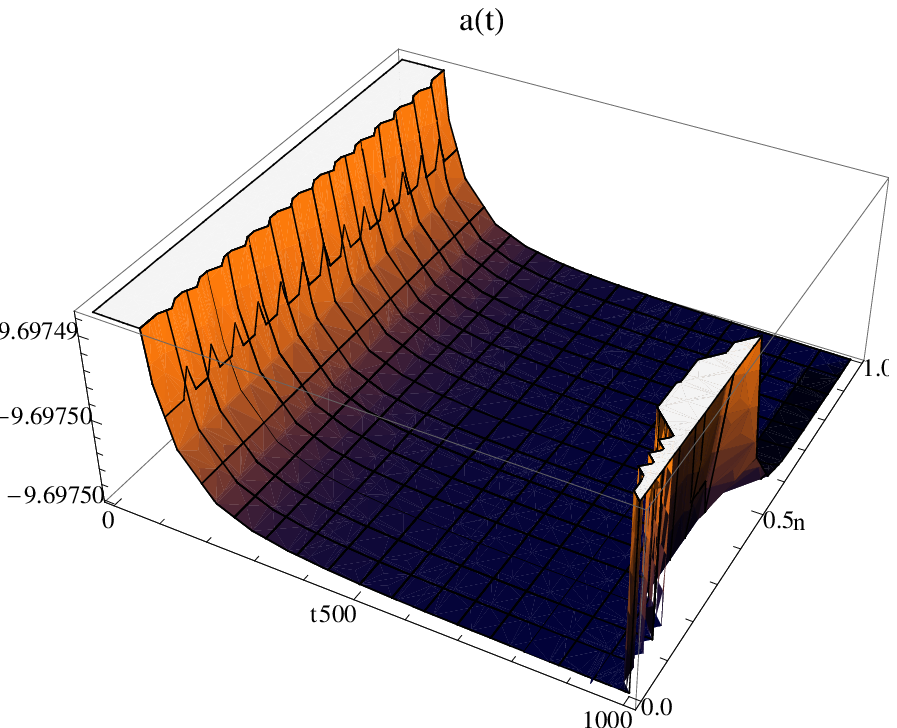}\hspace{1cm}
  \includegraphics[width=2in]{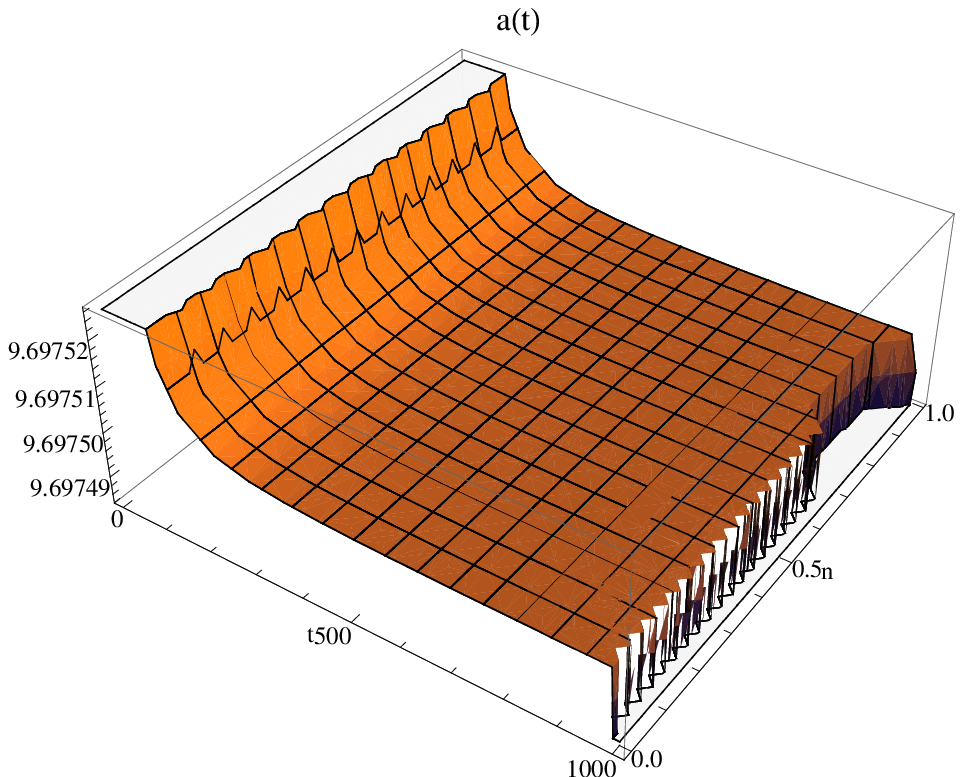}\hspace{0.1cm}
  \includegraphics[width=2in]{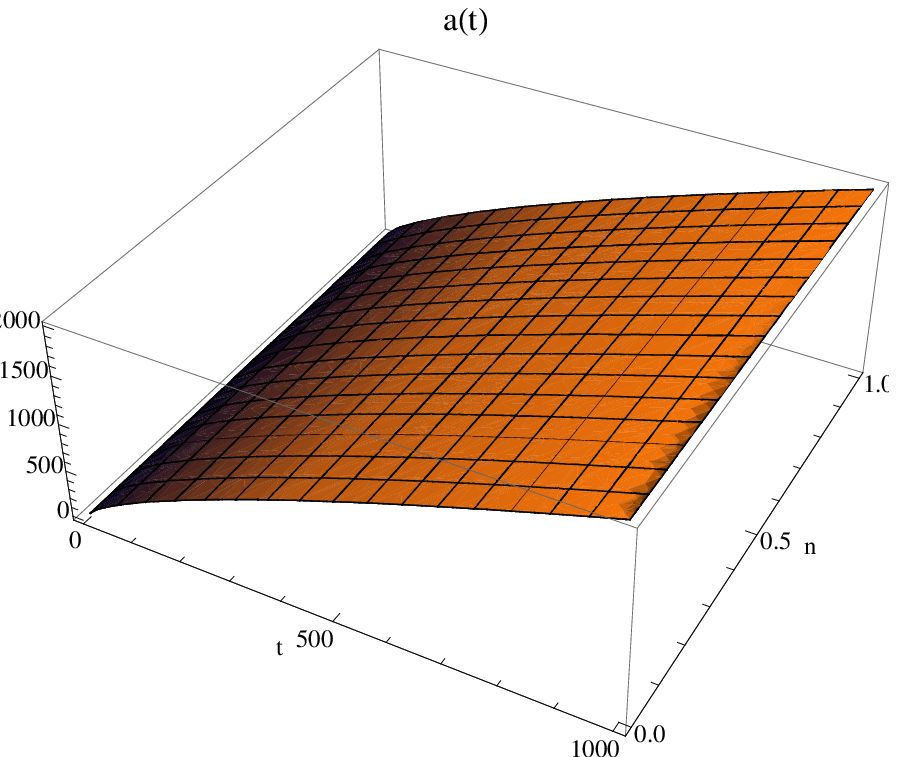}\hspace{0.7cm}
  \includegraphics[width=2in]{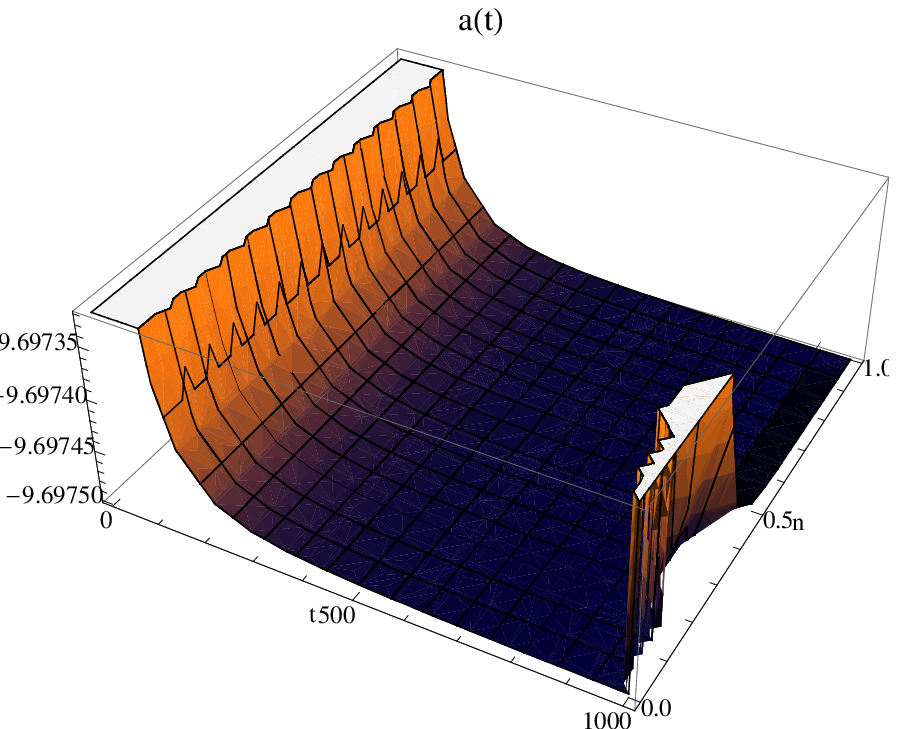}\hspace{0.7cm}
  \includegraphics[width=2in]{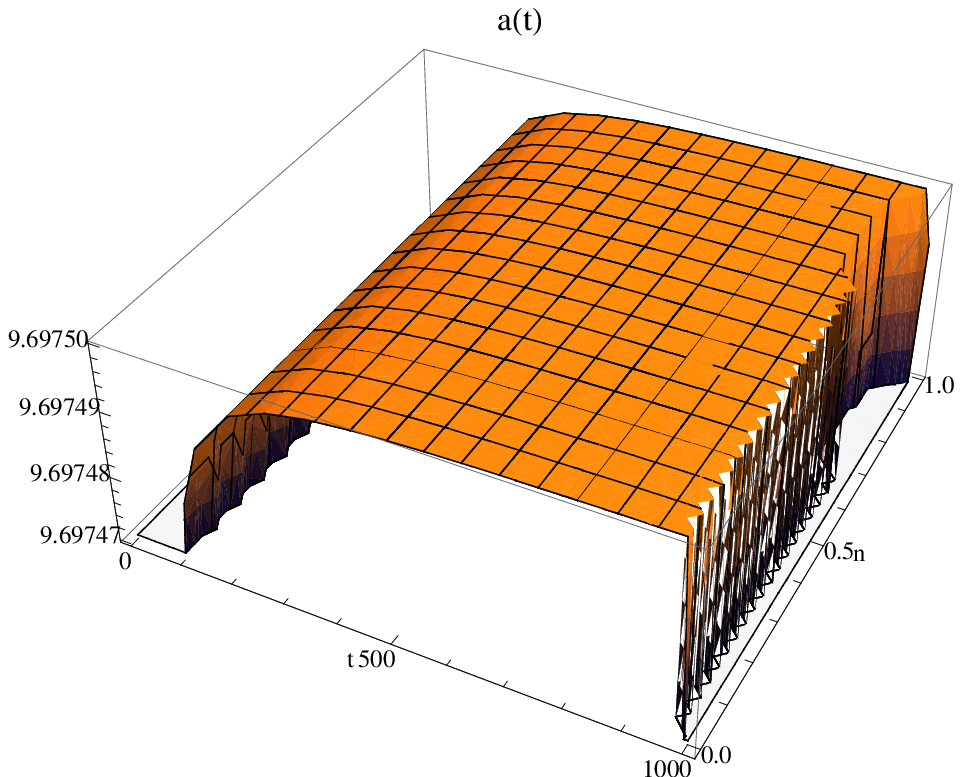}\hspace{0.7cm}
  \caption{The scale factor $a(t)$, density $\rho(t)$ and pressure $P(t)$ evolution with two typical values  $q=0.2$ and $q=0.8$ in the upper and lower plots, respectively. We have fixed $a_{s}=2000$, $M_{*}=M_{g}=m^{2}=1$, $t_{s}=1000$, $\beta_{0}=9.9$, $\beta_{1}=0$, $\beta_{2}=0.3$, $\beta_{3}=0$, $\beta_{4}=4$ and $\kappa=-1$.}
  \label{stable}
\end{figure*}

\textbf{Class II:}~~$\textbf{t}\rightarrow \textbf{t}_{s}$ \textbf{with} $n=1$
\textbf{and} $\textbf{q}\in \left(\textbf{0}\textbf{,}\textbf{1}\right]$ \\

In this class we have

\be
\label{Fbi45}
a(t_{s})\rightarrow a_{s},~~~ \dot{a}(t_{s})\rightarrow \dot{a}_{s}>0,~~~H(t_{s})\rightarrow H_{s} >0,~~~\ddot{a}(t_{s})\rightarrow \ddot{a}_{s}\geq 0,
\ee

\be
\label{Fbi46}
\rho (t_{s})\rightarrow \rho_{s}<0,~~~|P(t_{s})|\rightarrow P_{s}.
\ee

This result represents no future singularity. We have plotted the evolution of the quantities $a(t)$, $\rho(t)$ and $P(t)$ in Fig. $6$, for two typical values of $q$. We show that increasing $q$ value results in that $\rho(t)$ and $P(t)$ increase more rapidly for $t$ $\rightarrow$ $t_{s}$.\\\\

\begin{figure*}[ht]
  \centering
  \includegraphics[width=2in]{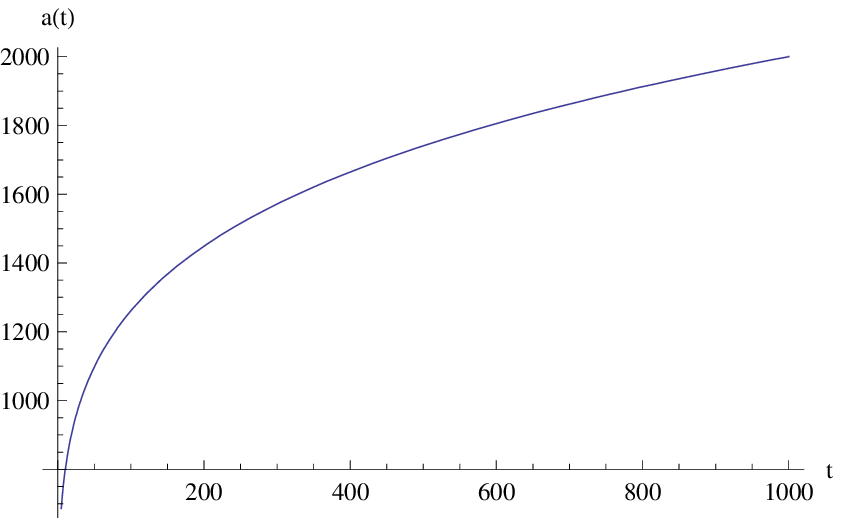}\hspace{1cm}
  \includegraphics[width=2in]{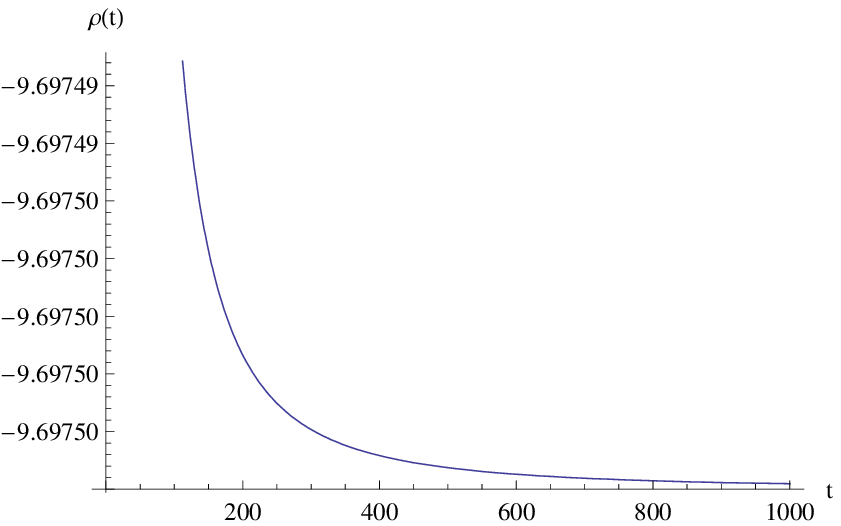}\hspace{1cm}
  \includegraphics[width=2in]{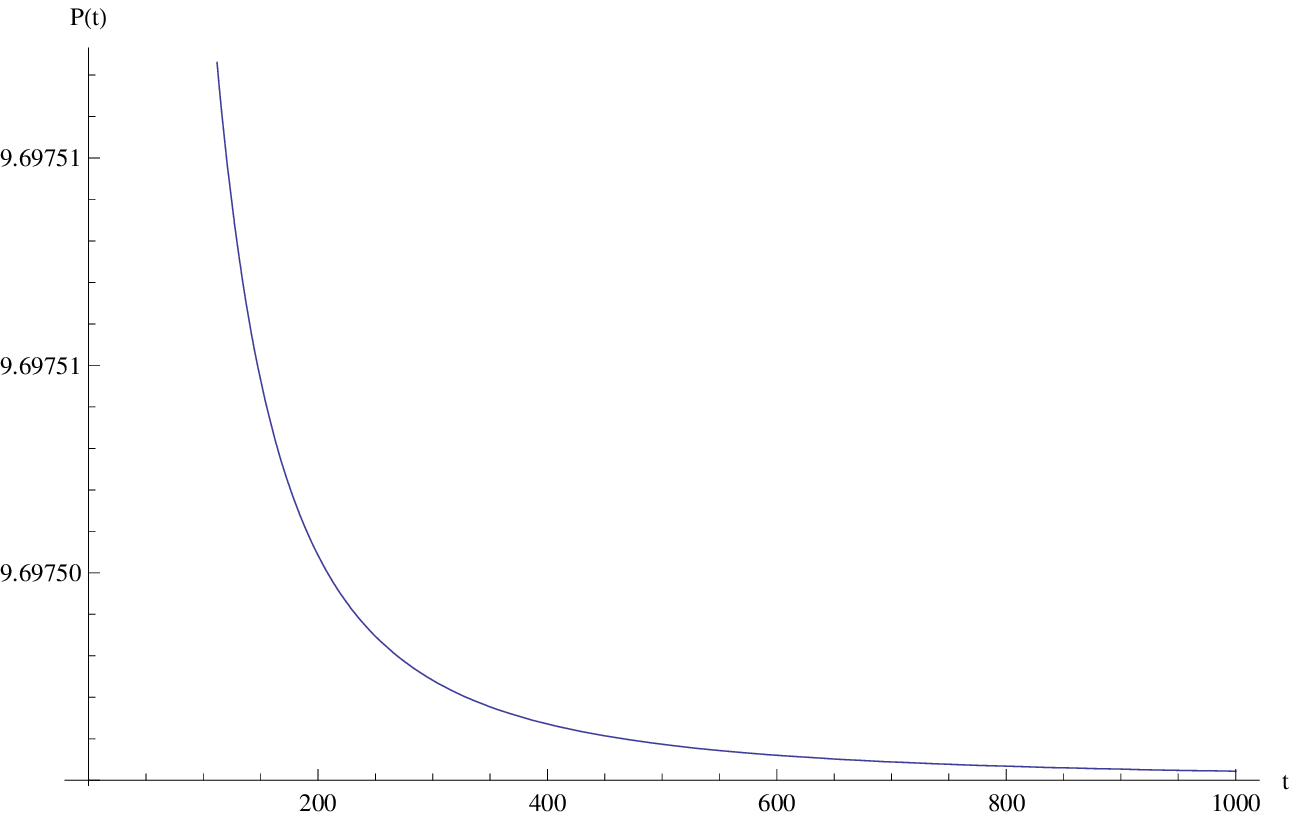}\hspace{0.1cm}
  \includegraphics[width=2in]{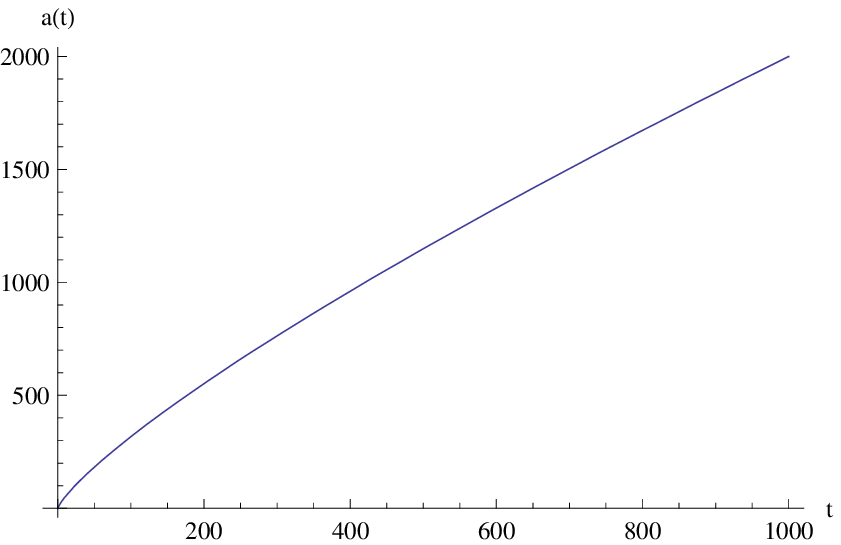}\hspace{0.7cm}
  \includegraphics[width=2in]{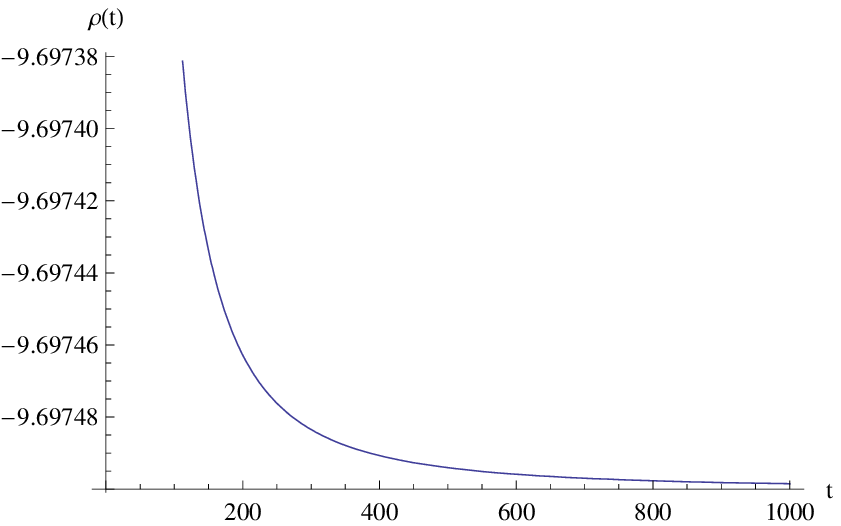}\hspace{0.7cm}
  \includegraphics[width=2in]{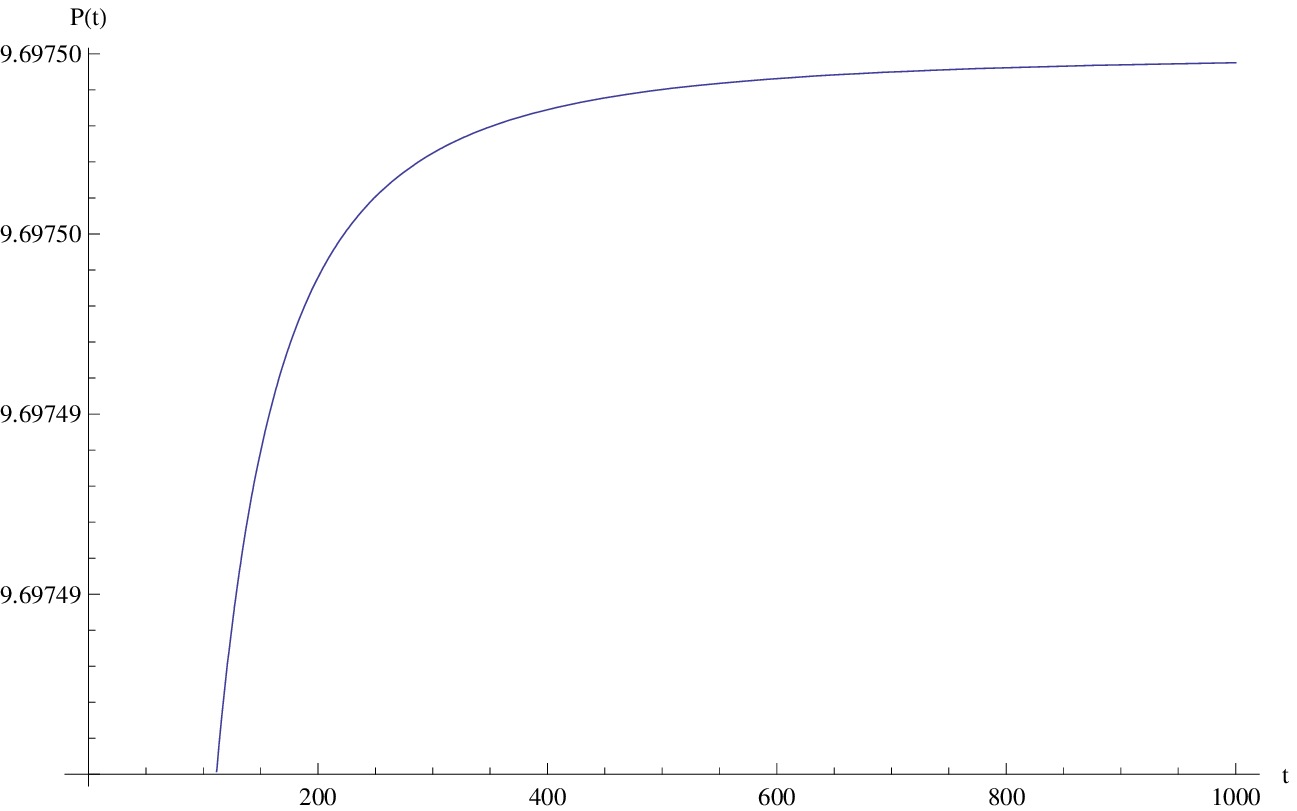}\hspace{0.7cm}
  \caption{The scale factor $a(t)$, density $\rho(t)$ and pressure $P(t)$ evolution with two typical values  $q=0.2$ and $q=0.8$ in the upper and lower plots, respectively. We have fixed $a_{s}=2000$, $M_{*}=M_{g}=m^{2}=1$, $t_{s}=1000$, $\beta_{0}=9.9$, $\beta_{1}=0$, $\beta_{2}=0.3$, $\beta_{3}=0$, $\beta_{4}=4$ and $\kappa=-1$.}
  \label{stable}
\end{figure*}

\textbf{Class III:}~~$\textbf{t}\rightarrow \textbf{t}_{s}$ \textbf{with} $\textbf{n}\in \left(\textbf{1}\textbf{,}\textbf{2}\right)$
\textbf{and} $\textbf{q}\in \left(\textbf{0}\textbf{,}\textbf{1}\right]$ \\

This class is identified by

\be
\label{Fbi47}
a(t_{s})\rightarrow a_{s},~~~ \dot{a}(t_{s})\rightarrow \dot{a}_{s}>0,~~~H(t_{s})\rightarrow H_{s} >0,~~~\ddot{a}(t_{s})\rightarrow \ddot{a}_{s}\geq 0,
\ee

\be
\label{Fbi48}
\rho (t_{s})\rightarrow \rho_{s}<0,~~~|P(t_{s})|\rightarrow \infty.
\ee

This result represents ``sudden'' singularity. The evolution of the quantities $a(t)$, $\rho(t)$ and $P(t)$ have been plotted in Fig. $7$, for two typical values of $q$. The figure implies that increasing $q$ value results in that $\rho$ and $P$ increase more rapidly for $t$ $\rightarrow$ $t_{s}$. In this singularity we can find the root of the pressure singularity in the relation (\ref{Fbi33}) in which we can see that the dynamical parameter $\dot{Y}$ is responsible for the infinity pressure at a finite time. If $\dot{Y}$ goes to infinity then according to (\ref{Fbi42}) we have $X(t)\rightarrow \infty$. In order to avoid this singularity we just can again refer to (\ref{Fbi33}) and suppose that $\beta_{_{2}}\gamma$ takes very small value so

\be
\label{Fbi48d}
\beta_{_{2}}\gamma\frac{\dot{Y}}{\dot{a}}\rightarrow~~\emph{\rm finite}.
\ee
Note that in the limit $\beta_{_{2}}\gamma\rightarrow 0$ the cosmological equations in the minimal massive bigravity approach to the Friedmann equations, but it seems that in this limit the asymptotic behavior of the minimal massive bigravity is dissimilar from GR at least in the sudden cosmological singularity case.

\begin{figure*}[ht]
  \centering
  \includegraphics[width=2in]{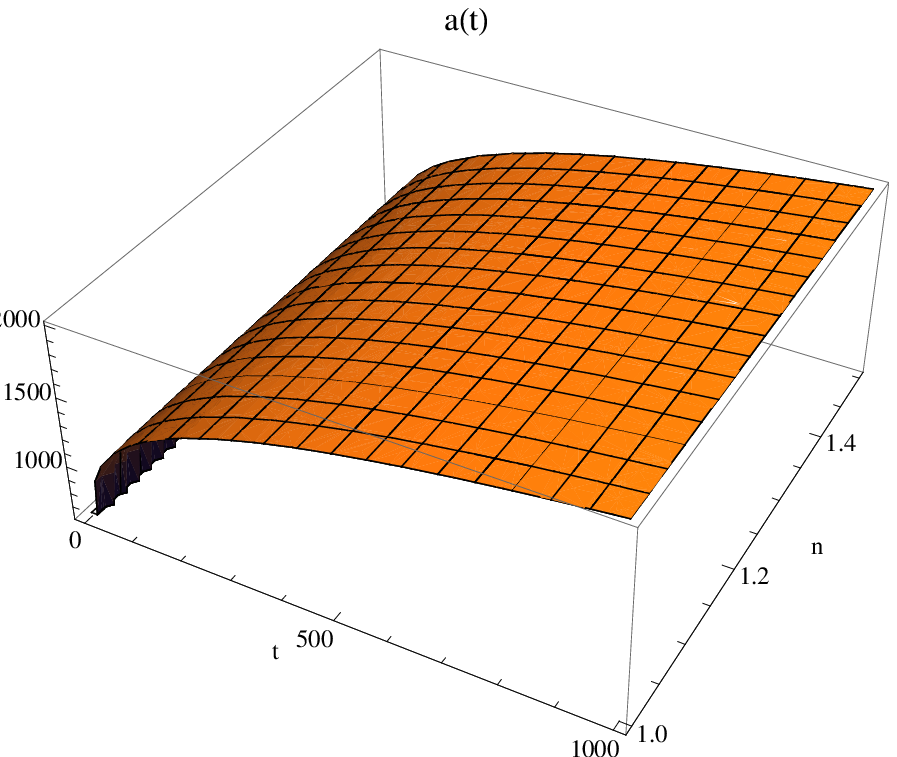}\hspace{1cm}
  \includegraphics[width=2in]{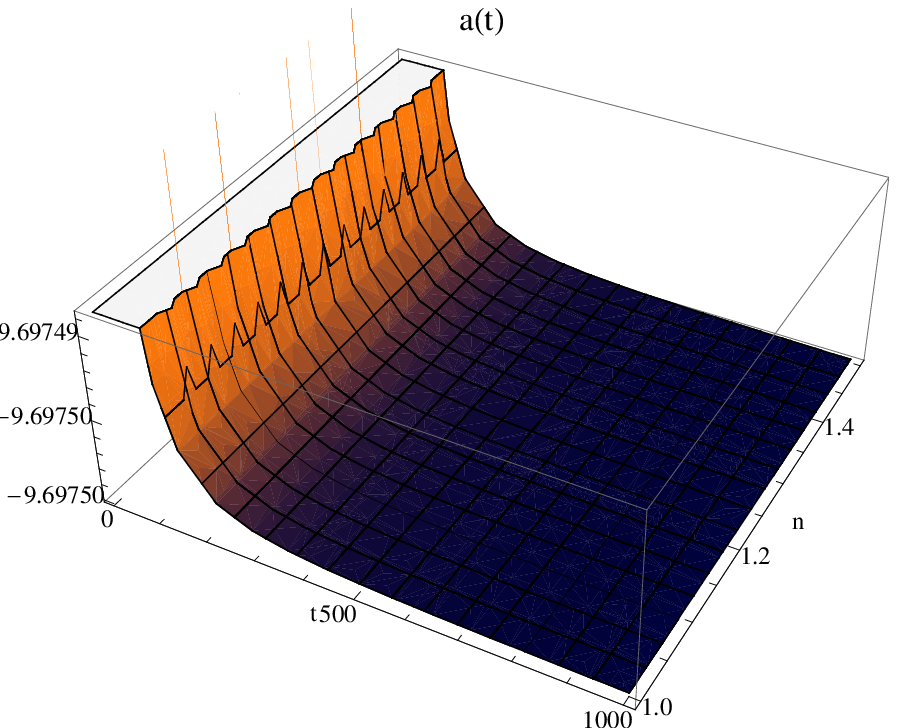}\hspace{1cm}
  \includegraphics[width=2in]{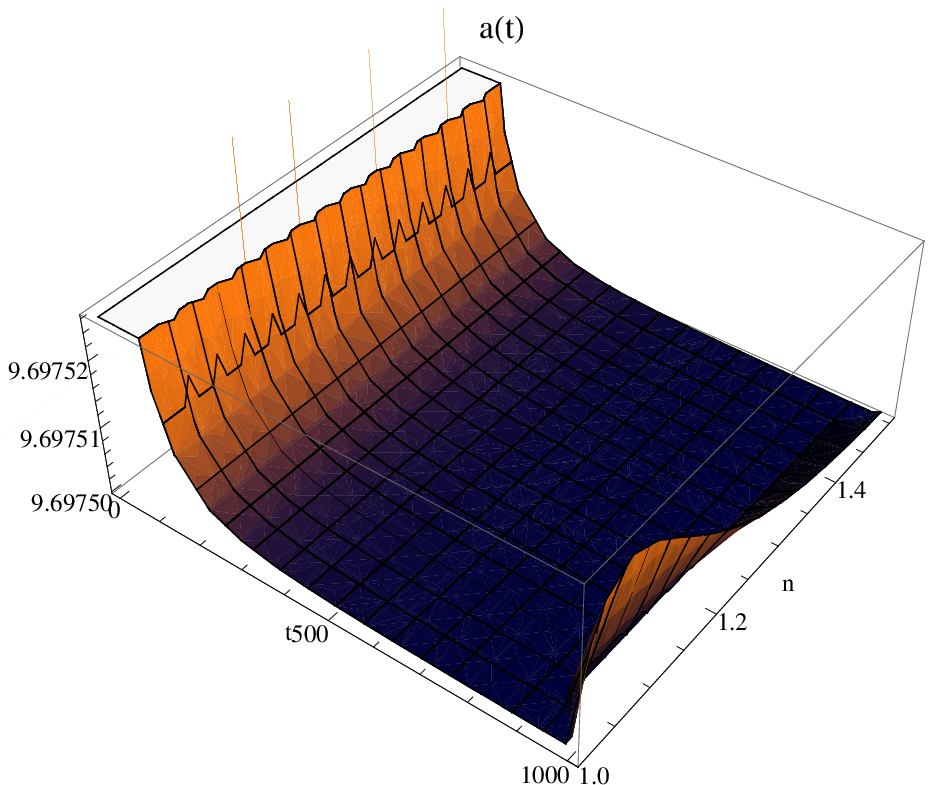}\hspace{0.1cm}
  \includegraphics[width=2in]{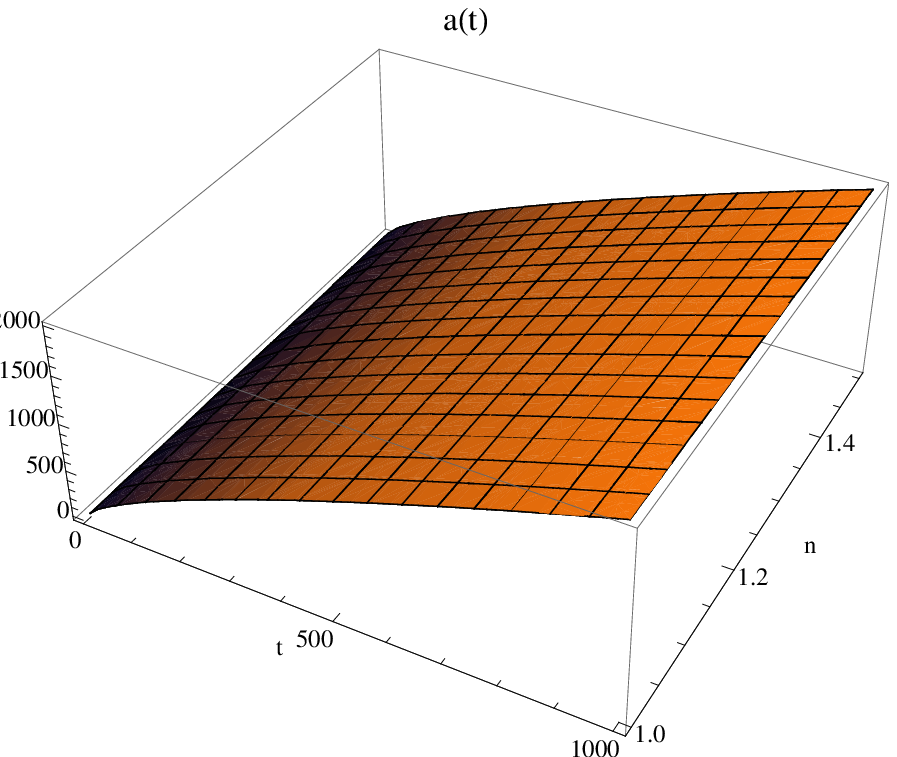}\hspace{0.7cm}
  \includegraphics[width=2in]{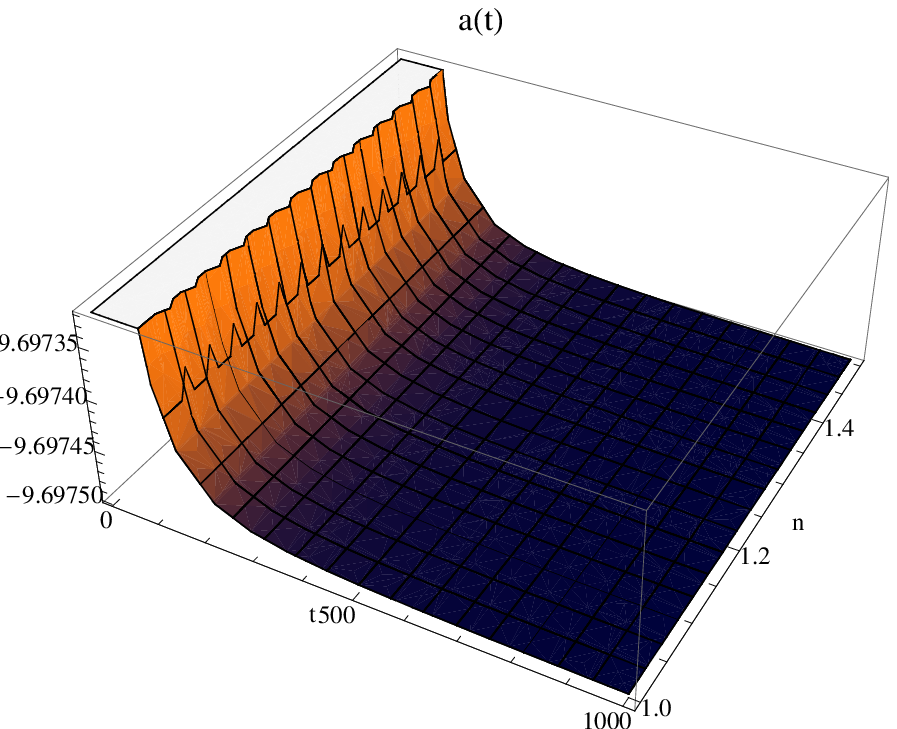}\hspace{0.7cm}
  \includegraphics[width=2in]{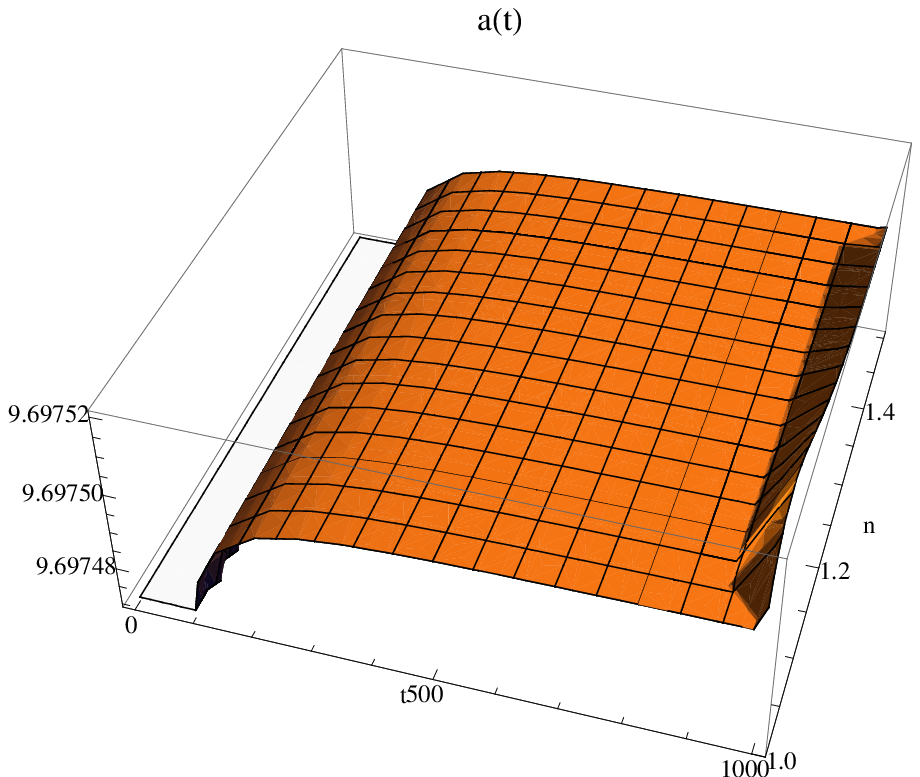}\hspace{0.7cm}
  \caption{The scale factor $a(t)$, density $\rho(t)$ and pressure $P(t)$ evolution with two typical values  $q=0.2$ and $q=0.8$ in the upper and lower plots, respectively. We have fixed $a_{s}=2000$, $M_{*}=M_{g}=m^{2}=1$, $t_{s}=1000$, $\beta_{0}=9.9$, $\beta_{1}=0$, $\beta_{2}=0.3$, $\beta_{3}=0$, $\beta_{4}=4$ and $\kappa=-1$.}
  \label{stable}
\end{figure*}

\textbf{Class IV:}~~$\textbf{t}\rightarrow \textbf{t}_{s}$ \textbf{with} $\textbf{n}\in \left[\textbf{2}\textbf{,}\infty\right)$
\textbf{and} $\textbf{q}\in \left(\textbf{0}\textbf{,}\textbf{1}\right]$ \\

In this class we face with

\be
\label{Fbi49}
a(t_{s})\rightarrow a_{s},~~~ \dot{a}(t_{s})\rightarrow \dot{a}_{s}>0,~~~H(t_{s})\rightarrow H_{s} >0,~~~\ddot{a}(t_{s})\rightarrow \ddot{a}_{s}\geq 0,
\ee

\be
\label{Fbi50}
\rho (t_{s})\rightarrow \rho_{s}<0,~~~|P(t_{s})|\rightarrow P_{s}.
\ee

This result represents no future singularity. The evolution of the quantities $a(t)$, $\rho(t)$ and $P(t)$ have been plotted in Fig. $8$, for two typical values of $q=0.2$ and $q=0.8$. The figure implies that increasing $q$ value results in that $\rho(t)$ and $P(t)$ increase more rapidly for $t$ $\rightarrow$ $t_{s}$.\\\\

\begin{figure*}[ht]
  \centering
  \includegraphics[width=2in]{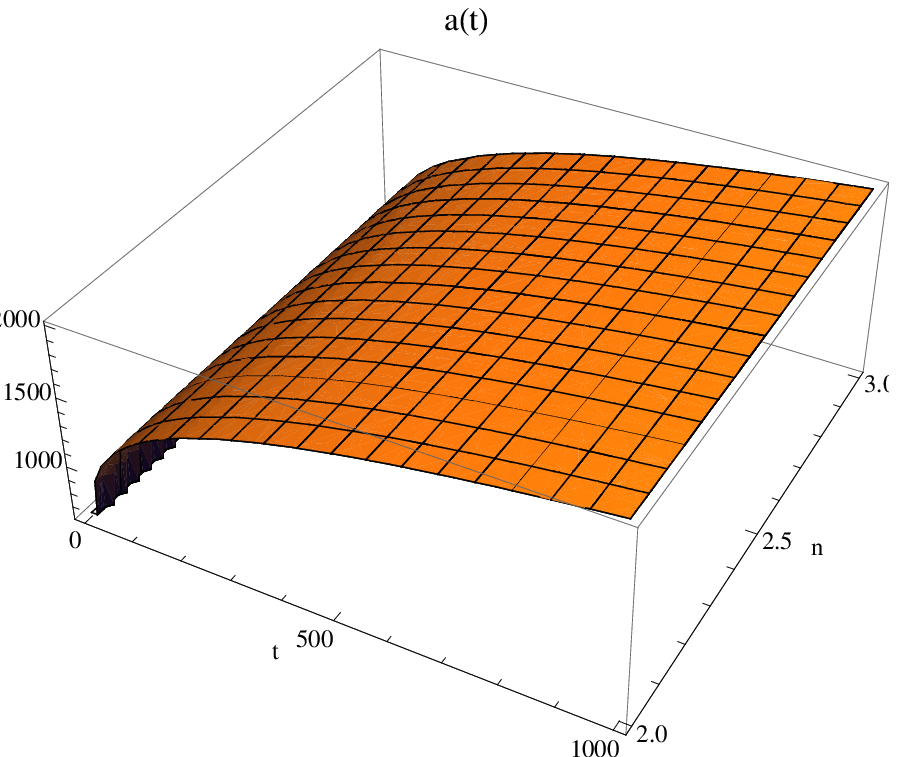}\hspace{1cm}
  \includegraphics[width=2in]{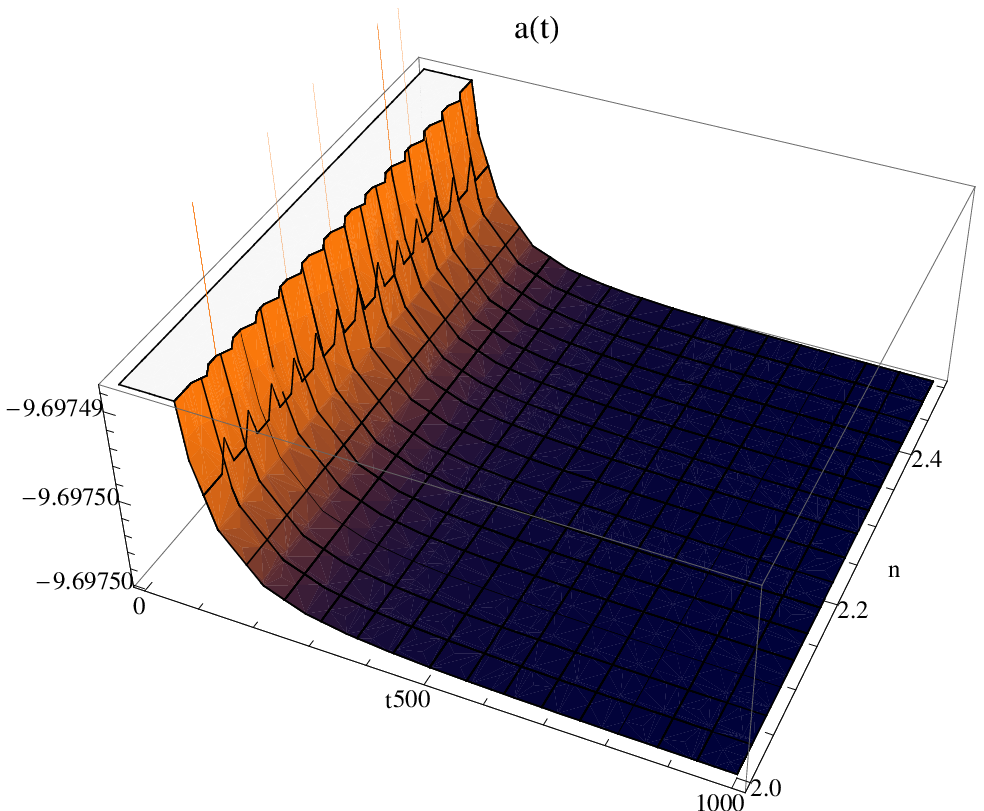}\hspace{1cm}
  \includegraphics[width=2in]{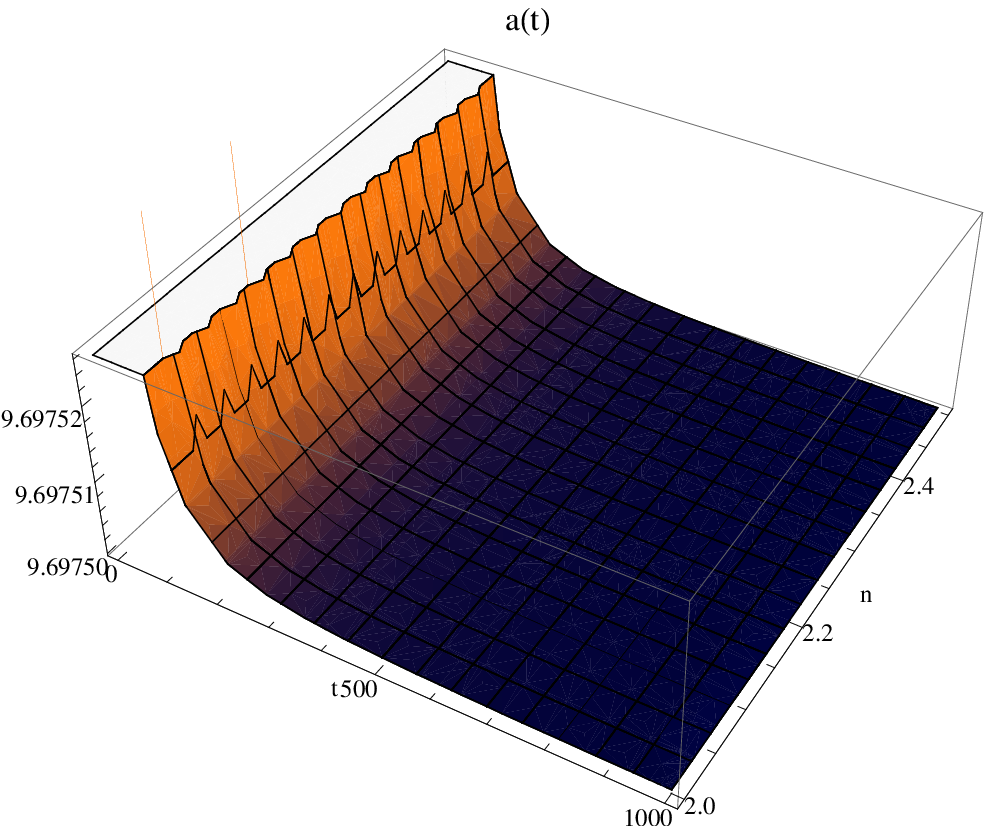}\hspace{0.1cm}
  \includegraphics[width=2in]{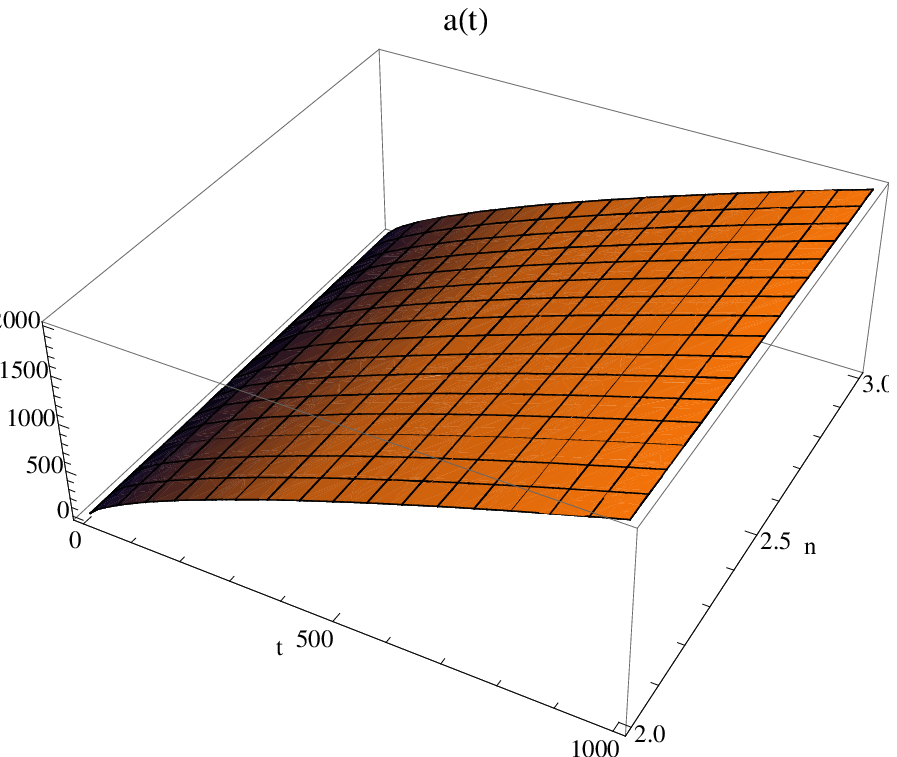}\hspace{0.7cm}
  \includegraphics[width=2in]{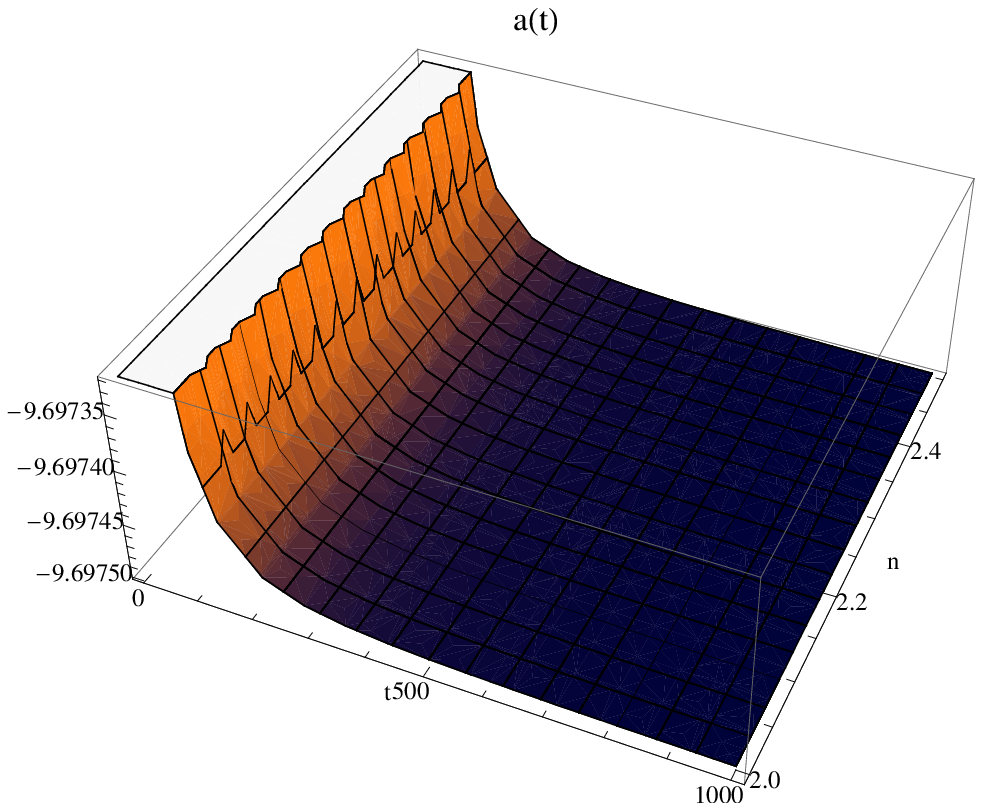}\hspace{0.7cm}
  \includegraphics[width=2in]{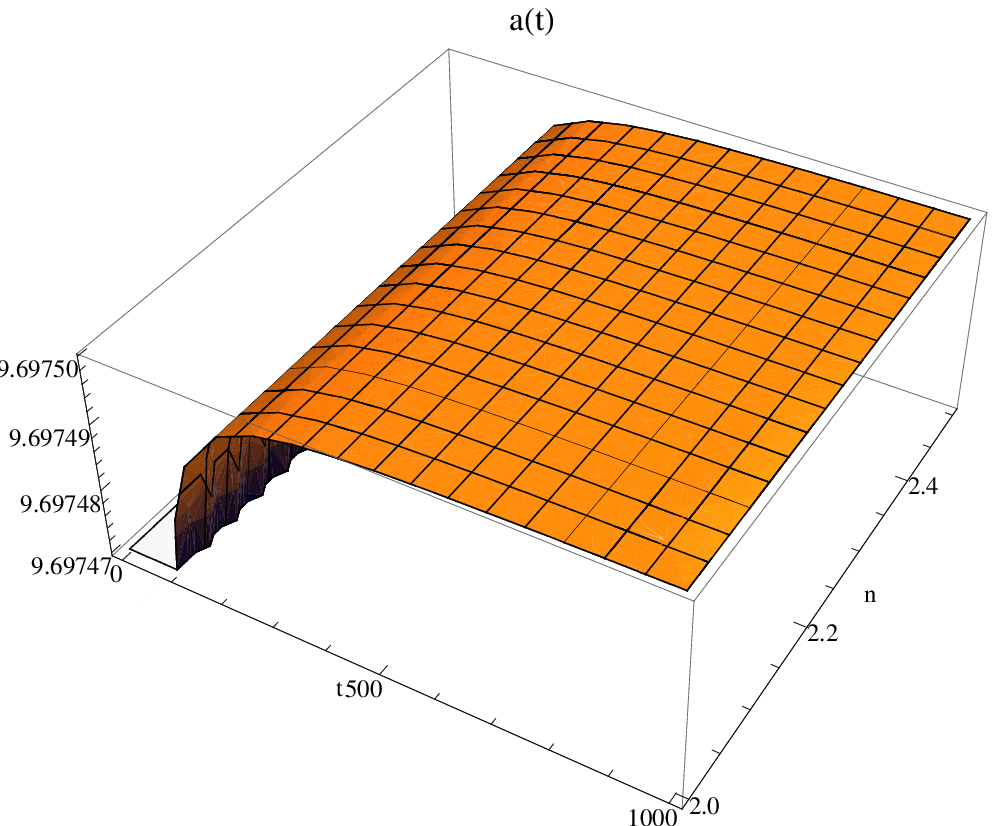}\hspace{0.7cm}
  \caption{The scale factor $a(t)$, density $\rho(t)$ and pressure $P(t)$ evolution with two typical values  $q=0.2$ and $q=0.8$ in the upper and lower plots, respectively. We have set $a_{s}=2000$, $M_{*}=M_{g}=m^{2}=1$, $t_{s}=1000$, $\beta_{0}=9.9$, $\beta_{1}=0$, $\beta_{2}=0.3$, $\beta_{3}=0$, $\beta_{4}=4$ and $\kappa=-1$.}
  \label{stable}
\end{figure*}

\section{Conclusions\label{Sec6}}
In this paper, we have considered the future cosmological singularities in the ghost-free massive gravity and massive bigravity models. By regarding the types of future singularities, we have studied the possible classes of finite-time future cosmological singularities for sudden, big rip, big freeze and big brake singularities.
In an expanding open Fridmann universe (which was mentioned in the introduction that people could have found open cosmological solutions in dRGT model), we have obtained the ``sudden singularity'' in which the physical quantities $\dot{a}(t)$, $H(t)$, $\ddot{a}(t)$ and $P(t)$ all go to infinity and cause such a singularity in a  finite-time future accompanying a finite fluid positive density and we can not eliminate these extracted singularities by no means of our theory parameters such as graviton mass because there is no coupling between for instance $m$ and dynamical quantities like $H(t)$ and $\ddot{a}(t)$,
nonetheless the singularities such as sudden finite-time future singularity that pressure approaches to infinity can basically be avoidable by taking a realistic equation of state for matter fields $P(\rho)$ by which the pressure profile is upper bounded by some well-defined density function. As a result, we can eliminate the pressure singularity at a finite-time in the future whereas the strong energy condition holds, see the first reference in \cite{28}.\\
Also, to complete our study in the finite-time future cosmological singularities we have focused on the massive bigravity theory and reviewed the modified Friedmann equations. In particular, for more simplicity we have considered the minimal massive bigravity model with $\beta_{1}=\beta_{3}=0$. We have found that there exists the possibility of occurring the sudden and big brake singularities. The sudden singularity can be eliminated in the minimal massive bigravity theory by taking $\beta_{_{2}}\gamma X(t)\rightarrow~\emph{\rm finite}$, nevertheless the theory approaches to GR regime.

In order to compare the finite-time future singularities in massive gravity and massive bigravity theories with cosmological singularities in GR, we can say that the results of our studding are similar to GR in which the sudden singularity arises and in fact it was predictable because the modification terms in the massive theories are just able to moderate the singularity not able to eliminate it and thus the sudden singularity in massive gravity is common whit GR. The distinction is just about the big brake singularity and also the behavior of the expanding scale factor in the comparison of massive gravity and massive bigravity theories and finally in the massive bigravity theories sudden singularity can be avoided contrary to massive gravity and GR.


\end{document}